\documentclass[11pt,letterpaper]{article}
\pdfoutput=1
\usepackage{jheppub}

\usepackage{amsmath}
\usepackage{amssymb}
\usepackage{epsfig,multicol,bbm}
\usepackage{graphicx}
\usepackage{tabu}
\usepackage{bm}
\usepackage{multirow}
\usepackage{bbold}
\usepackage[mathscr]{euscript}
\usepackage{makecell}

\newcommand{\TODO}[1]{}
\newcommand{\DK}[1]{}

\renewcommand{\d}[0]{\textrm{ d}}

\newcommand{\be}[0]{\begin{equation}}
\newcommand{\ee}[0]{\end{equation}}

\newcommand{\eq}[1]{Eq.~\eqref{eq:#1}}

\newcommand{\bea}{\begin{eqnarray}}
\newcommand{\eea}{\end{eqnarray}}

\newcommand{\nn}{\nonumber}

\newcommand{\eps}{\epsilon}

\DeclareMathOperator{\Tr}{Tr}

\newcommand{\CF}{C_F}
\newcommand{\TF}{T_F}

\newcommand{\as}{\alpha_s}

\newcommand{\cB}{\mathcal{B}}

\newcommand{\lb}{\Big{\lbrack}}
\newcommand{\rb}{\Big{\rbrack}}
\newcommand{\lp}{\Big{(}}
\newcommand{\rp}{\Big{)}}
\newcommand{\lbc}{\Big{\lbrace}}
\newcommand{\rbc}{\Big{\rbrace}}
\newcommand{\veb}[1]{#1_{T}^{\mathrm{cut}}}

\newcommand{\prp}[1]{#1_{\perp}}
\newcommand{\tra}[1]{#1_{T}}

\newcommand{\overbar}[1]{\mkern 1.5mu\overline{\mkern-1.5mu#1\mkern-0.3mu}\mkern 1.5mu}
\newcommand{\bigv}{\Big{\vert}}
\newcommand{\lngl}{\Big{\langle}}
\newcommand{\rngl}{\Big{\rangle}}

\newcommand{\etacut}{\eta_{\text{cut}}}
\newcommand{\mupf}{\mu^{\text{pf}}}
\newcommand{\nupf}{\nu^{\text{pf}}}
\newcommand{\ymax}{y_{\text{max}}}
\newcommand{\qmax}{Q_{\text{max}}}
\newcommand{\qmin}{Q_{\text{min}}}

\title{From Underlying Event Sensitive To Insensitive: Factorization and Resummation}

\author[a]{Daekyoung Kang, }
\author[b]{Yiannis Makris, }
\author[c]{and Thomas Mehen}

\affiliation[a]{Key Laboratory of Nuclear Physics and Ion-beam Application (MOE) and Institute of Modern Physics, Fudan University, Shanghai, China 200433}
\affiliation[b]{Theoretical Division T-2, Los Alamos National Laboratory, Los Alamos, NM, 87545}
\affiliation[c]{Department of Physics, Duke University, Durham, NC 27708}

\emailAdd{dkang@fudan.edu.cn}
\emailAdd{yiannis@lanl.gov}
\emailAdd{mehen@phy.duke.edu}

\abstract{In this paper we study the transverse energy spectrum for the  Drell-Yan process. The transverse energy is measured within the central region defined by a (pseudo-) rapidity cutoff. Soft-collinear effective theory (SCET) is used to factorize the cross section and resum large logarithms of the rapidity cutoff and ratios of widely separated scales that appear in the fixed order result. We develop a framework which can smoothly interpolate between various regions of the spectrum and eventually match onto the fixed order result. This way a reliable calculation is  obtained for the contribution of the initial state radiation to the measurement. By comparing our result for Drell-Yan against \textsc{Pythia} we obtain a simple model that describes the contribution from multiparton  interactions (MPI). A model with little or no dependence on the primary process gives results in agreement with the simulation. Based on this observation we propose MPI insensitive measurements. These observables  are insensitive to the MPI contributions as implemented in \textsc{Pythia} and we compare against the purely perturbative result obtained with the standard collinear factorization.}

\preprint{LA-UR-18-21306}

\begin{document}
\maketitle
\noindent


\section{Introduction}
\label{sec:intro}

Modern experimental and theoretical studies of processes in hadron colliders are often limited by our understanding of the underlying event which describes all that is seen by the detectors  that does not come directly from the primary hard process. In hadronic collisions understanding the various contributions to the underlying event  is crucial not only for testing quantum chromodynamics (QCD) but also in searches for new physics and precision measurements. 

The bulk of underlying event activity comes from multiparton interactions (interactions between the proton remnants from the hard process) and initial and final state radiation (ISR and FSR).  Although the contribution to the underlying event  from initial and final state radiation can be calculated in perturbation theory, contributions from multiparton interactions are more challenging to estimate. Currently the most effective way for integrating MPI with the hard process and partonic initial and final state showers are through models implemented in Monte Carlo  simulations. 

In experimental and Monte Carlo studies a class of observables known as MPI sensitive observables are used to probe the underling event activity in hadronic colliders. Transverse  energy, $E_T$, is such an observable and is defined as
\begin{equation}
  E_T(\etacut) = \sum_i p_T^{(i)} \Theta (\etacut - \vert \eta^{(i)} \vert)
\,,  \end{equation}
where $p_{T}^{(i)}$ is the scalar transverse momentum of particle $i$ and $\eta^{(i)}$ is its pseudo-rapidity.  For the CMS and ATLAS experiments at the large hadron collider (LHC) the cutoff parameter, $\etacut$, is typically chosen to be $\sim 2 - 2.5$ (see, for example,  
Refs.~\cite{CMS:2016etb,Chatrchyan:2012tb,Aad:2016ria,Aaboud:2017fwp,Aad:2014hia}). Other examples of such observables are the beam thrust~\cite{Stewart:2009yx,Stewart:2010pd,Berger:2010xi,Aad:2016ria} and the transverse thrust~\cite{Banfi:2010xy} but in this paper we will focus  on transverse energy.

MPI sensitive observables take large contributions from spectator-spectator interactions and it was shown in Refs.~\cite{Rothstein:2016bsq,Gaunt:2014ska,Zeng:2015iba} that these contributions are related to the violation of the traditional factorization due to Glauber gluon exchanges. In this paper we do not attempt to prove a factorization formula but we rather adopt an alternative approach where we include multiparton interactions through a model function convolved with the perturbative calculation from the collinear and soft factorization. We study the dependence of the model on the hard scale of the process using \textsc{Pythia} simulations  and we find that (for the LHC) below the TeV scale the MPI distribution is independent of the hard scale. The same result was found in Ref.~\cite{Papaefstathiou:2010bw} using {\ttfamily Herwig++} by studying different primary  processes (Higgs, $Z$, and $W^{\pm}$ production). The effect of MPI in Higgs transverse energy distributions was also studied in Ref.~\cite{Grazzini:2014uha}.

In Ref.~\cite{Hornig:2017pud} it was  shown that the factorization of the cross section depends on the region of phase-space under study, even for relatively large rapidity cutoff. Particularly, two regions of phase-space are identified,
\begin{align}
  \label{eq:regions}
  \mathrm{Region \; I:}\;\;\; &Q r \ll E_T \ll Q \nn \\
  \mathrm{Region \; II:}\;\;\; &E_T \lesssim Q r  \ll Q\; ,
\end{align}
where $r=\exp(-\etacut)$ is the cutoff ``radius'' and $Q$ the partonic center-of-mass energy. In this work we review the analysis of Ref.~\cite{Hornig:2017pud} and we illustrate how within the framework of soft-collinear effective theory~\cite{Bauer:2000ew,Bauer:2000yr,Bauer:2001ct,Bauer:2001yt} (SCET) we can study the effects of rapidity cutoff on resummed transverse energy distributions measured in hadronic collisions. We use the factorization of Ref.~\cite{Hornig:2017pud} and demonstrate that in the limit $E_T \gg Qr$ and with the appropriate choice of dynamical scales, this factorization reduces to the one introduced in Refs.~\cite{Papaefstathiou:2010bw,Tackmann:2012bt} for global measurements of transverse energy. In this limit the cross section is independent of the rapidity cutoff up to power corrections of $\mathcal{O}(Qr/E_T)$. To simplify the discussion we focus on the Drell-Yan process $pp \to \gamma^*( \to \ell^+ \ell^-) +X$, where the measurement of transverse energy is imposed on $X$. 

In region I since the cross section is independent of the rapidity cutoff, logarithmic enhancements from non-global effects are not important. However, in region II such effects are expected to become important for $E_T \ll Qr$, we find that for the values of $r$ we are interested resummation of global logarithms alone is sufficient to describe  transverse energy distribution   where it has significant support.

Since our formalism allows us to calculate the transverse energy spectrum for a wide range of  the rapidity cutoff parameter, it can be used for understanding the rapidity dependence of the MPI.  For example we found using \textsc{Pythia} that the mean transverse energy from MPI increases linearly with $\etacut$ for $1.5 <\etacut < 3.5$.

Relying on the observation that the model function is insensitive to the hard scale in  process  we propose an observable defined by MPI-sensitive transverse energy but designed to be MPI-insensitive such that we can make predictions for this observable using the standard soft and collinear factorization formula.  We refer to this observable as the subtracted transverse energy and is defined as the difference of the mean transverse energy for two different hard scales,
\begin{equation}
  \Delta E_T(Q,Q_0) \equiv \langle E_T(Q) \rangle - \langle E_T(Q_0) \rangle
  \,.
  \end{equation}
Comparing measurements of this observable against our analytic calculations we can determine if the assumptions  made in order to build the model are reasonable. We demonstrate that  this observable is independent of MPI contributions for phenomenologically relevant regions, when MPI are calculated using \textsc{Pythia}. Measurement of the mean traverse energy as a function of the hard scale was already performed for various processes in Refs.~\cite{Aad:2014hia,Aaboud:2017fwp}. This subtraction method can be generalized to other  additive quantities and as an example, we show that for beam thrust~\cite{Tackmann:2012bt,Stewart:2010qs,Stewart:2010pd,Banfi:2010xy} also is insensitive to MPI effects as generated by \textsc{Pythia}.


The factorization of the cross section for regions I and II (see Eq.(\ref{eq:regions})) within SCET is  discussed in Sections~\ref{sec:RI} and~\ref{sec:RII}, respectively and in Section~\ref{sec:merging} we discuss the merging of the corresponding factorizations with the use of profile scales. We describe the matching onto the fixed order result in QCD in Section~\ref{sec:QCD}. Furthermore in Section~\ref{sec:QCD} we give the assumptions made on the contribution of MPI which leads to the convolution of the perturbative result and a model function for the form of the true cross section. Including the MPI contribution using  \textsc{Pythia} we construct  a model function for the MPI that gives an accurate description of the the simulation data. The model we construct is independent of the partonic invariant mass.  Based on the assumptions that lead to the convolutional form of the cross section we introduce an observable insensitive to MPI in Section~\ref{sec:subtraction}. We confirm that these observables are MPI independent by comparing our purely  perturbative results to \textsc{Pythia} simulations. We conclude in Section~\ref{sec:conclusions}.
\section{Factorization}
\label{sec:fact}
In this section we illustrate  how within the framework of SCET we can reliably describe  the transverse energy distribution for the process $q \bar{q} \to \gamma^* +X$ for phenomenologically interesting  values of the transverse energy and the rapidity cutoff.  It was shown in Ref.~\cite{Hornig:2017pud} that when a rapidity cutoff is imposed the transverse energy distribution is insensitive to the cutoff parameter only in the region $E_T \gg Qr$. In this region the transverse energy spectrum can be described with the factorization theorem for the global case and for this reason we begin in section~\ref{sec:RI}  presenting a factorization theorem for  the global definition of $E_T$. In section~\ref{sec:RII}  we review the factorization of the cross section for $E_T \lesssim Qr$ and in section~\ref{sec:merging} we show how both regions can be described in a single factorization theorem with appropriate  choice of dynamical scales which we refer to as profile scales. Finally we discuss the matching onto the fixed order QCD result which describes the region $E_T \sim Q$ in section~\ref{sec:QCD}

 
 \subsection{Transverse energy as a global observable: region I}
 \label{sec:RI}

Here the transverse energy is defined as a global observable by,
\begin{equation}
  \label{eq:ETI}
E_T = \sum_{i\notin \{\ell^+,\ell^- \}} p_T^{(i)}\;,
\end{equation}
where $i$ extends over all the particles in the event other than the di-lepton  pair from the decay of the virtual photon. In the region where $E_T$ is parametrically smaller than the invariant mass of the di-lepton pair, the relevant modes to the measurement are the soft and collinear modes and their corresponding scaling is, 
\begin{align}
    \mathrm{soft:}\;\;\; p^{\mu}_{s} &= (p_s^{+},p_s^-,p_s^{\perp})\sim (E_T,E_T,E_T) \nn \\
  \mathrm{collinear:}\;\;\; p^{\mu}_{c} &= (p_c^{+},p_c^-,p_c^{\perp})\sim (E_T^2/Q,Q,E_T) \;,
\end{align}
where $p^{\pm}$ and $p^{\perp}$ are the light-cone and perpendicular components of momenta with respect to the beam axis. 
The effective field theory  that describes the dynamics and interactions of these modes is SCET$_{\text{II}}$. The hard scaling modes, $p^{\mu}_h\sim(Q,Q,Q)$, have been integrated out during the construction of the effective theory. The cross section can then be factorized into hard, soft, and collinear functions~\cite{Papaefstathiou:2010bw,Tackmann:2012bt}: 
\begin{equation}
  \label{eq:factI}
 \frac{d\sigma^{\text{(G)}}}{dy dQ^2 d\tra{E}}  = \sigma_0   H(Q;\mu)\times S_s(\tra{E};\mu,\nu) \otimes\cB_{q/P}^{\text{ G}}(x_1, E_T;\mu,\nu) \otimes \cB_{\overbar{q}/P}^{\text{ G}}(x_2,E_T;\mu,\nu)
 \,,
\end{equation}
where $Q$ and $y$ are the invariant mass and rapidity of the virtual photon and the parton momentum fraction is given by $x_{1,2}=Q e^{\pm y}/\sqrt{s}$. The hard process, $q\bar{q} \to \gamma^*(\to \ell^+ \ell^-) +X$ is described through the hard function $H$, which is the the product of matching coefficients from matching QCD onto SCET. The initial state radiation (ISR) from soft and collinear emissions is incorporated  within the soft, $S_s$, and beam functions, $\cB_{q/P}$, respectively. In addition, the beam functions contain information regarding the extraction of a parton, $a$, from the proton. The operator definition of the beam function is~\cite{Fleming:2006cd,Stewart:2009yx},
\begin{equation}
  \label{eq:beamOD}
  \mathcal{B}_{q/P}^{\text{ G}}(x_B,\tra{E},\mu)=\Tr \lb\sum_{X} \delta(\tra{E}-\tra{E}^X)\lngl P_n(k) \bigv \bar{\chi}_n(0) \frac{\gamma^- }{2} \bigv X \rngl \lngl X \bigv \delta (p^- -\overbar{\mathcal{P}}_n) \chi_n(0) \bigv P_n(k) \rngl \rb,
\end{equation}
where $P_n(k)$ is the proton with momentum $ k^{\mu} = (0^+,k^-,\prp{0})$, and  $x_B=p^-/k^-$ is the fraction of the proton momentum carried by the quark field.  Although the beam function is a non-perturbative object for $E_T \gg \Lambda_{\text{QCD}}$ it can be matched onto the (also non-perturbative but well known) collinear parton distribution functions (PDFs). This is achieved through a convolution of perturbative calculable matching coefficients and the PDFs evaluated at a common scale, $\mu$,~\cite{Stewart:2009yx}:
\begin{equation}
  \cB^{\text{ G}}_{j/P}(x_B,E_T;\mu,\nu) = \sum_i \int_{x_B}^{1} \frac{dx}{x} \mathcal{I}^{\text{ G}}_{j/i}(x,E_T;\mu,\nu) f_{i/P}\lp \frac{x_B}{x};\mu \rp
  \end{equation}
where $\mathcal{I}^{\text{ G}}_{j/i}$ are the matching coefficients and we use the superscript G  to denote that these are the matching coefficients for the case of global measurement in contrast to the case where rapidity cutoff is implemented. We analyze the latter case in the following section.  The next-to-leading order (NLO) matching coefficients are given in Appendix~\ref{ap:A}. The soft function can be calculated order by order in perturbation theory using the operator definition and the NLO result is given in Ref.~\cite{Tackmann:2012bt}.  The Born cross section, $\sigma_0$ is defined by
\begin{equation}
\label{eq:sigma0}
\sigma_0  \equiv\frac{16\pi^2\alpha_{EM}^2 e_q^2}{3 N_{c}   Q^2 E_{\text{cm}}^2}.
\end{equation}

All elements of Eq.(\ref{eq:factI}) depend on the  factorization scale, $\mu$,  and thus need to be evaluated at a common scale before combining them to construct a scale independent cross section. For this reason we use renormalization group (RG) methods that allow us to evolve each function from its canonical scale up to an arbitrary scale. This will result in a transverse energy distribution with resummed logarithms of ratios of $E_{T}$ and $Q$, up to a particular logarithmic accuracy. In this work we will study the next-to-leading logarithmic prime (NLL') accuracy.

The perturbative expansions of the beam matching coefficients and the soft function suffer from rapidity divergences that are not regulated with pure dimensional regularization. For this reason we use the rapidity regulator introduced in Refs.~\cite{Chiu:2012ir,Chiu:2011qc}. Although the rapidity regulator dependence cancels at the level of cross section, the rapidity scale, $\nu$, introduced during the regularization procedure allows us to resum the complete set of logarithms of $E_T/Q$. It is only after solving the rapidity-renormalization-group (RRG) equations that we may resum all logarithms of $E_T/Q$ up to a particular accuracy. Thus the final result for the resummed distribution is
\begin{multline}
  \label{eq:finalG}
  \frac{d\sigma^{\text{(G)}}}{dy dQ^2 d\tra{E}} = \sigma_0\; \mathcal{U}_H(\mu_{ss},\mu_{H}) H(Q,\mu_H) \times \mathcal{V}_{ss} (E_T;\mu_{ss},\nu_{ss},\nu_{\cB}) \otimes S_s(E_T;\mu_{ss},\nu_{ss})  \\
   \otimes  \cB^{\text{ G}}_{q/P}(x_a,E_T;\mu_{\cB},\nu_{\cB}) \otimes  \cB^{\text{ G}}_{\bar{q}/P}(x_{\bar{a}},E_T;\mu_{\cB},\nu_{\cB})
  \end{multline}
where $\mathcal{U_H}$ and $\mathcal{V}_{ss}$ are defined in Appendix~\ref{ap:C} as  the solutions to the following RG and RRG equations,
\begin{align}
  \label{eq:RGE}
  \frac{d}{d\ln \mu} H(Q;\mu) &= \gamma_{\mu}^{H} (Q,\mu)H(Q;\mu) ,\nn \\
  \frac{d}{d\ln \nu} S_s(E_T;\mu,\nu) &= \gamma_{\nu}^{ss}(E_T,\mu) \otimes  S_s(E_T;\mu,\nu)\;,
\end{align}
where
\begin{equation}
  [g\otimes f](E_T) \equiv \int dE_T'\; f(E_T-E_T') g(E_T')\;.
  \end{equation}
 More details regarding the RG and RRG properties  of the transverse energy or broadening  dependent functions can be found in Refs.~\cite{Tackmann:2012bt,Chiu:2012ir}. The canonical scales $\mu_H, \mu_{ss}$, and $\mu_{\cB}$  are used as the initial conditions for the solutions of the differential equations in Eqs.(\ref{eq:RGE}) and are chosen such that they minimize the logarithms in the perturbative expansion of the corresponding functions:
\begin{align}
  \mu_H &= Q\;, & \mu_{ss} &= \mu_{\cB}= E_T.
  \end{align}
Similarly for the rapidity scales we have,
\begin{align}
  \nu_{ss} &= E_T\;, & \nu_{\cB} &= Q. 
\end{align}

As mentioned earlier in Ref.~\cite{Hornig:2017pud} it was shown that  the the cross section in region I is well described by the global factorization. That means,
\begin{equation}
   \frac{d\sigma^{\text{(I)}}}{dy dQ^2 d\tra{E}} \simeq  \frac{d\sigma^{\text{(G)}}}{dy dQ^2 d\tra{E}}\;.
\end{equation}
We use the above equation to describe the spectrum in region I and later in section~\ref{sec:merging} to show that we can describe both regions I and II with a single factorization theorem.


\subsection{Transverse energy with rapidity cutoff: region II}
\label{sec:RII}

The transverse energy with rapidity cutoff is defined by,
\begin{equation}
  \label{eq:ETII}
  E_T(\etacut) = \sum_{i\notin \{\ell^+,\ell^- \}} p_T^{(i)} \;\Theta(\etacut - \vert \eta^{(i)} \vert)
\,,\end{equation}
where $\eta^{(i)}$ is pseudo-rapidity of the $i$-th  particle. As in the global case, we sum over all the particles  in the event excluding the di-lepton pair and $\etacut$ is the cutoff parameter\footnote{Also for simplicity of notation, for the rest of the paper we omit the dependence on $\etacut$ in $E_T(\etacut)$ and we specify in the text when we refer to the global definition from Eq.(\ref{eq:ETI}). }.

Region II, $\tra{E} \lesssim Q r$, is discussed in detail in Ref.~\cite{Hornig:2017pud}. Here we summarize only the main results necessary  for the analysis relevant to this work. As was illustrated in Ref.~\cite{Hornig:2017pud}, in this region we can identify an additional soft scale which is collinear enough to resolve the boundary of the rapidity cutoff. This mode was first introduced in Ref.~\cite{Chien:2015cka}\footnote{See also Ref.~\cite{Bauer:2011uc} for similar extensions of SCET and the collinear-soft modes.} in the context of jet-radius resummation. Adopting the naming scheme of Ref.~\cite{Chien:2015cka} we refer to this mode as soft-collinear. Thus all the relevant modes are: (u-)soft, collinear, and soft-collinear. The corresponding scaling is,
\begin{align}
  \mathrm{(u-)soft:}\;\;\; p^{\mu}_{s} &\sim (E_T,E_T,E_T) \nn \\
  \mathrm{soft-collinear:}\;\;\; p^{\mu}_{sc} &\sim (E_T r,E_T/r,E_T) \nn \\
  \mathrm{collinear:}\;\;\; p^{\mu}_{c} &\sim (Qr^2,Q, Qr) \;.
\end{align}
These collinear and soft-collinear modes are associated with the direction of one of the beams, similar modes exist for the direction of the other beam. The effective theory that describes these modes is SCET$_{++}$ and in this region the cross section factorizes in the following way, 
\begin{align}
  \label{eq:factII}
  \frac{d\sigma^{\text{(II)}}}{dy dQ^2 d\tra{E}}  &= \sigma_0   H(Q)\times S_s(\tra{E}) \otimes [S_{n} \otimes \cB_{q/P}^{\text{ II}}](x_1, E_T,r) \otimes [S_{\overbar{n}}\otimes \cB_{\overbar{q}/P}^{\text{ II}}](x_2,E_T, r) \nn \\ 
  &= \sigma_0   H(Q)\times S(\tra{E},r) \otimes\cB_{q/P}^{\text{ II}}(x_1, E_T,r) \otimes \cB_{\overbar{q}/P}^{\text{ II}}(x_2,E_T,
  r)
  \,,\end{align}
where $S_s(E_T)$ is the same global soft function that appears in Eq.(\ref{eq:factI}) and $S_n(E_T,r)$ is the soft-collinear function describing the contribution from soft-collinear modes near the cutoff boundary. In the first line we combined soft and soft-collinear functions into the total soft function, $S$,
\begin{equation}
\label{eq:SsSn}
  S(\tra{E}, r;\mu) = S_s (E_T;\mu,\nu) \otimes S_{n}  (E_T,r;\mu,\nu)\otimes S_{\overbar{n}} (E_T,r;\mu,\nu) .
\end{equation}
We note that the soft-collinear functions depend  on the rapidity scale $\nu$. This is due to the fact that, compared to the global case, the rapidity divergences (and thus the rapidity scale dependence) appear in the soft-collinear function rather than in the beam function. The beam functions are rapidity-finite and take, contributions from radiation within two distinct regions of phase-space, below the rapidity cutoff ($\etacut < \eta$) and beyond the cutoff ($\etacut > \eta$). Radiation below the cutoff  contributes only to the so-called unmeasured beam function which is proportional to $\delta(E_T)$ and contributes only to the zeroth bin of transverse energy. Radiation beyond the cutoff will contribute to the beam function through a power corrections of $\mathcal{O}(E_T/Qr)$. These power corrections could be ignored in the small transverse energy limit but are important in the regime where $E_T \sim Qr$.  Thus the beam function can be written as,
\begin{equation}
  \cB_{a/P}^{\text{ II}}(x,E_T,r;\mu) =  \cB_{q/P}(x, r;\mu) \delta(E_T) + \Delta \cB_{q/P} (x,E_T, r;\mu)
 \,.
\end{equation}
The beam function can be matched onto the collinear PDFs when $E_T \gg \Lambda_{\text{QCD}}$,
\begin{equation}
  \cB^{\text{ II}}_{j/P}(x,E_T,r;\mu) = \sum_i \int_{x}^{1} \frac{dz}{z} \mathcal{I}^{\text{ II}}_{j/i}(z,E_T,r;\mu) f_{i/P}\lp \frac{x}{z};\mu \rp
 \,, \end{equation}
where the matching coefficient $\mathcal{I}^{\text{ II}}_{a/i}$ can be written as,
\begin{equation}
  \label{eq:matcingII}
   \mathcal{I}^{\text{ II}}_{j/i}(x,E_T,r;\mu) =  \mathcal{I}_{j/i}(x,r;\mu)\delta(E_T) +\Delta B_{j/i}(x,E_T,r;\mu)
\end{equation}
The first term, $\mathcal{I}_{j/i}$, is the term that determines the unmeasured beam function. The perturbative expansion of this term contains UV divergences that need to be regulated and renormalized. This procedure determines the RG anomalous dimension and  evolution of the beam function, $\cB^{\text{ II}}_{j/P}$. The second term in Eq.(\ref{eq:matcingII}), $\Delta B_{j/i}$, gives the contribution to the power corrections that appear in the beam function.  This term requires zero-bin subtraction and is finite. The implicit dependence on the factorization scale $\mu$  in $\Delta B_{j/i}$ is due to the strong coupling constant. The operator definition of the beam function for region II  and the one-loop result for the corresponding matching coefficients  are given in Section 3 of Ref.~\cite{Hornig:2017pud}.

The resummed distribution involves evolving each term in the factorization theorem from its canonical scale to a common scale both in virtuality and rapidity. Similarly to the global case this is achieved through the solution of the corresponding RG equations. For the final result we have
    \begin{align}
  \label{eq:finalII}
  \frac{d\sigma^{\text{(II)}}}{dy dQ^2 d\tra{E}} = &\sigma_0\; \mathcal{U}_H(\mu_{ss},\mu_{H})\;  H(Q,\mu_H)  \; \mathcal{U}_{\cB}(\mu_{ss},\mu^{\text{II}}_{\cB}) \;  \mathcal{U}_{\cB}(\mu_{ss},\mu^{\text{II}}_{\overbar{\cB}}) \mathcal{V}_{ss} (E_T;\mu_{ss},\nu_{ss},\nu_{sc}) \nn \\ 
     & \otimes   S_s(E_T;\mu_{ss},\nu_{ss})
     \otimes  [S_n(E_T,r;\mu_{ss},\nu_{sc}) \otimes\cB^{\text{ II}}_{q/P}(x_a,E_T,r;\mu^{\text{II}}_{\cB})]  \nn \\[5pt]
     &\otimes  [ S_n(E_T,r;\mu_{ss},\nu_{sc})\otimes \cB^{\text{ II}}_{\bar{q}/P}(x_{\bar{a}},E_T,r;\mu^{\text{II}}_{\overbar{\cB}}) ]
\,, \end{align}
where the virtuality scales are
\begin{align}
  \mu_{H}&=Q\;, & \mu^{\text{II}}_{\cB/\overbar{\cB}} &= Q r e^{\pm y }\;, & \mu_{ss} &= \mu_{sc} = E_T\;.
\end{align}
We note that, in contrast to the global measurement, the two beams are evaluated at two distinct scales. For central events the two scales  are of the same order of magnitude but have different values depending on the rapidity of the virtual photon. This is a consequence of the rapidity cutoff since imposing such a constraint breaks boost invariance. This can be avoided by choosing a dynamic value of the cutoff parameter in a boost invariant way, i.e. $\etacut(y) = \etacut \pm y$.\footnote{We use $+$ for the beam direction, $n_{\mathcal{B}}$, and $-$ for the opposite direction, $n_{\mathcal{\overbar{B}}}= \bar{n}_{\mathcal{B}}$. } Our one loop results are then modified with the replacement $\etacut \to \etacut(y) $ and this gives us a boost invariant scale $\mu^{\text{II}}_{\cB/\overbar{\cB}} = Q e^{\etacut}$. With this choice we ensure that the jet scale is always parametrically smaller than the hard scale for all values  of the virtual photon's rapidity. Although a boost invariant definition of the rapidity cutoff is phenomenologically preferred, experimentally fixed cutoff is used and therefore here we proceed with the same choice. The rapidity  scales are,
\begin{align}
  \nu_{ss} &= E_T \;, & \nu_{sc} &= \frac{E_T}{ r}.
\end{align}

In the next section we discuss how modifying these scales and using the factorized cross section in Eq.(\ref{eq:finalII}) lead to a result that can describe both region I and II with a smooth interpolation in the intermediate regime.


\subsection{Profile scales and merging}
\label{sec:merging}
 
The goal of this section is to  show that in the limit $r\to0$ and  $\tra{E} \gg Qr$ the factorization for Region II (i.e. Eg.(\ref{eq:finalII})) matches onto that for the global measurment (i.e., Eq.(\ref{eq:finalG})) with the appropriate choice of dynamical scales  which we refer to as profile scales. That is,
\begin{align}
  d\sigma^{\text{II}} \Big{\vert}_{\text{pf}} &\xrightarrow{E_T \ll Qr}  d\sigma^{\text{II}}\;, & d\sigma^{\text{II}} \Big{\vert}_{\text{pf}} &\xrightarrow{E_T \gg Qr}  d\sigma^{\text{G}}
  \end{align}

The exact form of the the profile scales is not important but they need to satisfy the following asymptotic behavior,
\begin{align}
    \label{eq:profiles} 
 \mupf_{\overbar{\cB}}(E_T \ll Qr) &\sim \mu^{\text{II}}_{\overbar{\cB}}\;,&   \mupf_{\cB}(E_T \ll Qr) &\sim \mu^{\text{II}}_{\cB}\;,&   \nupf_{sc} (E_T \ll Qr) &\sim \nu_{sc}\;, \nn \\[5pt]
 \mupf_{\overbar{\cB}}(E_T \gg Qr) &\sim \mu_{\cB} = \mu_{ss}\;,&            \mupf_{\cB}(E_T \gg Qr) &\sim \mu_{\cB} = \mu_{ss}\;,&   \nupf_{sc} (E_T \gg Qr) &\sim \nu_{\cB}\;.
\end{align}
and
\begin{equation}
   d\sigma^{\text{II}} \Big{\vert}_{\text{pf}}  =  d\sigma^{\text{II}} \lp \mu^{\text{II}}_{\cB/\overbar{\cB}} \to  \mupf_{\cB/\overbar{\cB}},\;\;\mu_{sc} \to  \mupf_{sc} \rp
  \end{equation}
To see why this set of scales is appropriate for the matching of the two regimes consider evolution kernels that appear in  Eqs.(\ref{eq:finalG}) and (\ref{eq:finalII}) which are,
\begin{align}
\Pi^{\text{(G)}} (E_T;\{ \mu_i\} ) \equiv  &\; \mathcal{U}_H(\mu_{ss},\mu_H) \mathcal{V}_{ss} (E_T;\mu_{ss},\nu_{ss},\nu_{\cB}), \nn  \\[5pt]
\Pi^{\text{(II)}} (E_T;\{ \mu_i\} )  \equiv   &\; \mathcal{U}_H(\mu_{ss},\mu_H)  \; \mathcal{U}_{\cB}(\mu_{ss},\mu^{\text{II}}_{\cB}) \;  \mathcal{U}_{\cB}(\mu_{ss},\mu^{\text{II}}_{\overbar{\cB}}) \mathcal{V}_{ss} (E_T;\mu_{ss},\nu_{ss},\nu_{sc}) 
\end{align}
respectively. In region II where $E_T \ll Qr $, the transverse energy distribution should be described by $\d\sigma^{\text{(II)}}$ and thus the introduction of profile scales has no influence in the form of the factorization theorem. This is true since in that region the profiles reduce to the scales that they replace (see first line of Eq.(\ref{eq:profiles})). Therefore we have,
\begin{equation}
\Pi^{\text{(II)}} (E_T;\{ \mu_i\} ) \Big{\vert}_{\text{pf}} \xrightarrow{E_T \ll Qr} \Pi^{\text{(II)}} (E_T;\{ \mu_i\} )\;.
\end{equation}
 In the other region, $E_T \gg Qr$, the beam profiles equal the global soft scale, $\mu_{ss}=E_T$, thus the beam evolution kernels,  $\mathcal{U}_{\cB}$, reduce to the identity,
\begin{equation}
   \mathcal{U}_{\cB}(\mu_{ss},\mupf_{\cB}) \xrightarrow{E_T \gg Qr} \mathcal{U}_{\cB}(\mu_{ss},\mu_{ss}) = 1. 
  \end{equation}
Since the soft-collinear rapidity profile scale, $\nupf_{sc}$, is asymptotically reaching the beam rapidity scale for the global measurement, we have,
\begin{equation}
  \label{eq:Kmatching}
 \Pi^{\text{(II)}} (E_T;\{ \mu_i\} ) \Big{\vert}_{\text{pf}} \xrightarrow{E_T \gg Qr} \Pi^{\text{(G)}} (E_T;\{ \mu_i\} )\;.
  \end{equation}
For the rest of this section we demonstrate   that up to power corrections Eq.(\ref{eq:Kmatching}) can be extended to the NLL, NLL', and NNLL cross section. Since this has been shown for the evolution kernels we only need to show the same holds for the fixed order terms at $\mathcal{O}(\alpha_s^0)$ for NLL and $\mathcal{O}(\alpha_s^1)$ for the NLL' and NNLL cross-sections. At $\mathcal{O}(\alpha_s^0)$ this is trivial since both cases reduce to the Born cross section. At $\mathcal{O}(\alpha_s^1)$ we note that the hard function, $H(Q;\mu)$, and the global-soft function, $S_s(E_T;\mu,\nu)$, appear in both factorization theorems, therefore is sufficient to  show
\begin{equation}
  \label{eq:limit}
   S_n(E_T,r;\mu,\nu) \otimes\cB^{\text{ II}}_{a/P}(x,E_T,r;\mu)  \xrightarrow{E_T \gg Qr} \cB^{\text{ G}}_{a/P}(x,E_T;\mu,\nu) + \mathcal{O} \lp\frac{E_T}{Qr} \rp
\end{equation}
Since this task is more technical  we leave the details for Appendix~\ref{ap:B} and here we give a phase-space based argument using the corresponding operator definitions. For example  the operator definition of the soft-collinear function is~\cite{Chien:2015cka},
\begin{equation}
  S_n(E_T,r) = \frac{1}{N_C} \Tr \lb \sum_{X_{sc}}\delta(E_T - E_T^{X_{sc}}(\etacut)) \lngl 0 \bigv T [ V_{n}^{\dag} U_{n} ] \bigv X_{sc} \rngl \lngl X_{sc} \bigv T[U_{n}^{\dag} V_{n}] \bigv 0 \rngl \rb,
\end{equation}
which is proportional to $\delta(E_T - E_T^{X_{sc}}(\etacut))$ with $E_T^{X_{sc}}(\etacut)$ evaluated using Eq.(\ref{eq:ETII}). Taking the limit $r\to0$ (or equivalently $\etacut \to \infty$),
\begin{equation}
  \delta(E_T - E_T^{X_{sc}}(\etacut)) \xrightarrow{r \to 0} \delta(E_T - E_T^{X_{sc}}),
  \end{equation}
where on the r.h.s. $E_T^{X_{sc}}$  is defined globally.  The above equation can be understood in the following way: contributions to the regions of phase-space where particles are emitted within the cone are proportional to  the size of available phase-space volume to the power of the number of particles in the cone region (i.e., $(V_{\text{cone}})^{\text{\#-of particles in cone}}$). Thus, in the small cone limit these corners of phase-space will be suppressed  compared to the regions where all particles in $\vert X_{sc} \rangle$ are emitted within the measured region and may be ignored. This corresponds to a global definition of transverse energy.  Working in the $\overline{\text{MS}}$ scheme any higher order correction gives scaleless integrals and thus
\begin{equation}
  S_n(E_T,r) \xrightarrow{r \to 0}  S^{(0)}_n(E_T,r) =  \delta(E_T).
\end{equation}
This corresponds to no contribution to the measurement from soft-collinear modes, and suggests that in that limit the soft-collinear modes are redundant. This should be expected since if we define $z_{sc} \equiv p^-_{sc}/p_{c}^- \sim E^{X_{sc}}_T/(Qr)$ and demand $z_{sc} \ll 1$, then as we take the limit $r \to 0$ we unavoidably have $E^{X_{sc}}_T \to 0$. The same argument holds for the case of the beam function giving,
\begin{equation}
  \cB^{\text{ II}}_{a/P}(x,E_T,r;\mu) \xrightarrow{r \to 0}\cB^{\text{ G}}_{a/P}(x,E_T;\mu,\nu)\;,
\end{equation}
which up to  power corrections gives Eq.(\ref{eq:limit}) to all orders. To get the exact form of power correction we work with the cumulant functions at $\mathcal{O}(\alpha_s)$ in Appendix~\ref{ap:B}. The calculations of Appendix~\ref{ap:A} and~\ref{ap:B} confirm the claim that with a single factorization theorem and an appropriate choice of dynamical scales we can describe both regions of phase space. 


\subsection{Matching onto fixed order}
\label{sec:QCD}

In order to describes the transverse energy spectrum in the region $E_T \sim Q$ we need to match the resummed distribution to the fixed order (FO) result from the full theory. This is necessary in order to include power corrections of $E_T/Q$ not described by the effective theory. Furthermore, in this region logarithms of $E_T/Q$ are not large and thus the FO result correctly describes the transverse  energy spectrum. A smooth interpolation for the intermediate regime can be achieved by adding to the resummed distribution the difference of the full theory FO and effective theory FO result,
\begin{equation}
  \label{eq:matchingA}
  \frac{d\sigma}{d E_T} =  \frac{d\sigma^{\text{II}}}{d E_T}\Big{\vert}_{\text{pf}} + \lp \frac{d\sigma^{\text{FO}}}{d E_T}\Big{\vert}_{\mu=Q}- \frac{d\sigma^{\text{G, FO}}}{d E_T}\Big{\vert}_{\mu= Q}  \rp,
\end{equation}
where we obtain $\d\sigma/dE_T$ by integrating over $dy$ in the region $y\in(-\ymax,\ymax)$ and over $dQ$ in the region $Q\in(\qmin,\qmax)$, where $\qmax$, $\qmin$, and $\ymax$, define the kinematic cuts for the photon's rapidity and invariant mass.. In order to remain within the central region we need to impose $\ymax < \etacut$. In Eq.(\ref{eq:matchingA}) $d\sigma^{\text{FO}}/dE_T$ is the $\mathcal{O}(\alpha_s)$ full QCD result where no rapidity cutoff is imposed. Alternatively one can use,
\begin{equation}
  \label{eq:matchingB}
  \frac{d\sigma}{d E_T} =  \frac{d\sigma^{\text{II}}}{d E_T}\Big{\vert}_{\text{pf}} + \lp \frac{d\sigma^{\text{FO}}(\etacut)}{d E_T}\Big{\vert}_{\mu=Q}- \frac{d\sigma^{\text{II, FO}}}{d E_T}\Big{\vert}_{\mu= Q}  \rp,
\end{equation}
where now $d\sigma^{\text{FO}}(\etacut)$ is the full QCD result where the rapidity cutoff is imposed. The difference between Eq.(\ref{eq:matchingA}) and Eq,(\ref{eq:matchingB}) are power corrections which we already neglected during the construction of the factorization theorem. 

Note that $d\sigma^{\text{G, FO}}/dE_T$ does not depend on the rapidity scale $\nu$ since rapidity divergences cancel at fixed order in the convolution of the soft function and the beam functions. The $\mu$ scale dependence of  $d\sigma^{\text{G, FO}}/dE_T$ and $d\sigma^{\text{FO}}/dE_T$ comes from the running of the strong coupling and the scale dependence of PDFs. The choice of $\mu$ need to be the same for both so that detailed cancellation of the two is achieved in the region $E_T \ll Q$ where the resummed  distribution describes the spectrum. Detailed cancellation also needs to be achieved between $d\sigma^{\text{G, FO}}/dE_T$ and $d\sigma^{\text{II}}/dE_T$ in the region  $E_T \gtrsim Q$ where the fixed order result describes the spectrum. For this reason we need to turn-off evolution at $E_T \gtrsim Q$. This can be easily done choosing $\mu=Q$ and  using the profile scales in Eq.(\ref{eq:profiles}) and replacing $\mu_{ss},\mu_{\cB} \to \mu_{ss}^{\text{pf}}(E_T)$ and $\nu_{ss} \to \nu_{ss}^{\text{pf}}(E_T)$ where
\begin{align}
  \label{eq:profiles2}
  \mupf_{ss}(E_T < Q) &\sim E_T\;, &  \nupf_{ss}(E_T < Q) &\sim E_T \nn \;,\\
  \mupf_{ss}(E_T \gtrsim Q) &\sim \mu_H \sim Q \;, & \nupf_{ss}(E_T \gtrsim Q) &\sim \nupf_{sc}(Q) \sim Q. 
\end{align}
\begin{figure}[t!]
  \centerline{\includegraphics[width = \textwidth]{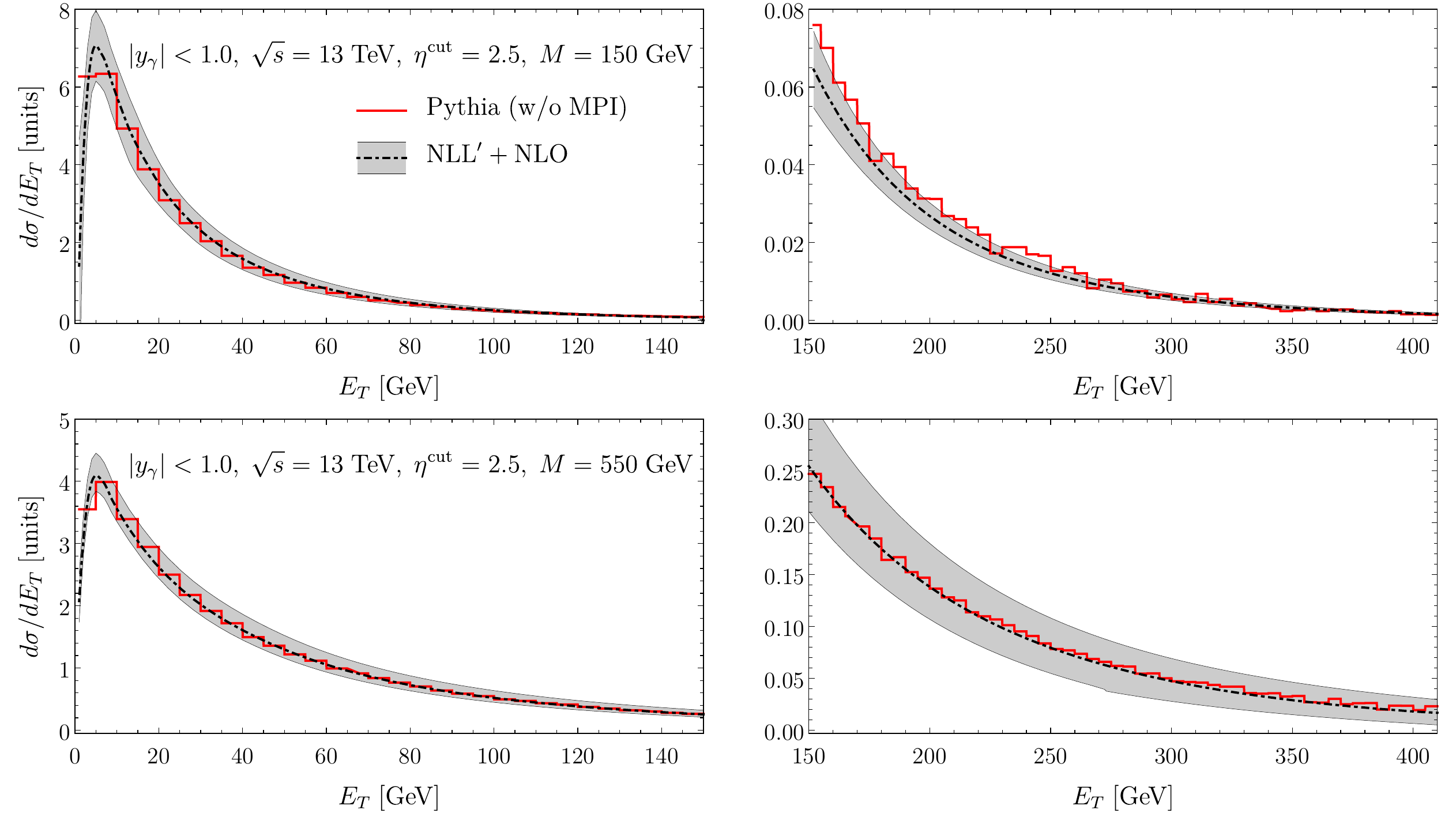}}
\caption{ The NLL-prime resumed distributions evaluated from Eq.(\ref{eq:matchingA})  (black dot-dashed) compared against the result from \textsc{Pythia} partonic distributions (red solid).}
\label{fig:al2p5}
\end{figure}

Comparing our matched NLL' result against a simulation using MadGraph~\cite{Alwall:2014hca}+\textsc{Pythia}~\cite{Sjostrand:2006za,Sjostrand:2007gs}, we find good agreement. The hard process $pp \to \gamma^* $ is performed in MadGraph and then showered by \textsc{Pythia}. We use \textsc{Pythia} build-in matrix element (ME) corrections for describing the distribution in the far tail. As discussed in Refs.~\cite{Sjostrand:2006za,Sjostrand:2007gs} this corresponds up to one additional hard emission from the initial state partons. This is sufficient for our case since we are matching only to NLO corrections in that region.\footnote{As discussed in the introduction, we will in Section~\ref{sec:subtraction} consider moments of this distribution, particularly the first moment. For phenomenological applications that require studies of higher moments, where contributions from the far tail are further enhanced, one needs to consider higher hard parton multiplicity. In Monte Carlo  simulations this is achieved through a ``Match and Merge'' procedure~\cite{Sjostrand:2006za,Sjostrand:2007gs}. Since here the simulation analysis is included only for purposes of comparison with the analytic NNL'+NLO result the default build in ME correction of \textsc{Pythia} are sufficient.}  In Figure~\ref{fig:al2p5}, we show the comparison in the peak region (left) and  tail region (right) for the choice  $\etacut = 2.5$.

The error band is estimated by varying all scales by a factor of two and one-half around their canonical values. The total error is calculated by adding in quadrature all variations. Caution is necessary here since the scales choice is implemented through the profile functions in order to transition from one region to the other. That requires that a single profile function will change with any of the scale variations in order to ensure the proper transition without double counting the variations. For example, the global-soft and beam profiles will change accordingly when we consider hard scale variation in order to freeze evolution in the far tail but should remain unchanged for $E_T \ll Q$. We collected all the details on the choice of profile functions and scale variation in Appendix~\ref{ap:profiles}.

In order to compare our analytic result with the partonic  distributions in \textsc{Pythia}  we turned off the multi-parton interactions and hadronization. The non-perturbative/hadronization effects on the resummed distributions  can be studied using the operator definition of the soft and collinear functions~\cite{Korchemsky:2000kp,Bauer:2002ie,Lee:2006fn,Lee:2007jr}. Usually the hadronization effects are included through a convolution of the soft function or the cross section with a model function (which needs to be determined from experiment). The convolution is  over  the measured observable and thus the  model function depends on the observable. The form of the model function is usually  determined using the operator product expansion to get the first few moments. This was done for various event and jet-shape observables such as  thrust, event-shape angularities, jet mass, groomed-jet mass, $D_2$, e.t.c.~\cite{Hornig:2009vb,Kang:2013lga,Kang:2018qra,Moult:2017okx,Stewart:2014nna}. Ref.~\cite{Becher:2013iya}, studies the non-perturbative effects in transverse momentum dependent (TMD) distributions and jet broadening in $e^+ e^-$ which are  most closely related to the measurement presented in this paper.

In contrast, contributions from multi-parton interactions (MPI) are not very well understood theoretically, and a systematic approach for describing these effects has yet to be developed. The subject of MPI and how our formalism can be used to study its effect is discussed in the next section.


\section{Multiparton interactions}
The origin of MPI is from secondary interactions of the beam remnants through Glauber exchanges. These interactions are known to break factorization in measurements of global observables but cancel in inclusive cross-sections. A variety of MPI sensitive observables are used in experimental studies for understanding the properties of underlying event (UE) but a comparison to the theory is currently impossible. In this paper we propose a prescription to describe MPI contributions to transverse energy with a rapidity cutoff. In experimental measurements of UE common choices for the rapidity cutoff parameter are  $\etacut = 2$ and $\etacut =2.5$ (for example, see Refs.~\cite{Aaboud:2017fwp,Aad:2016ria,Chatrchyan:2012tb,CMS:2016etb,Aad:2014hia}). Our prescription is based on the following two conjectures:
\begin{itemize}
\item{Contributions to underlying event from MPI can be modeled by a convolution of a model function with perturbative results, }
\item{The MPI model function is insensitive to hard scale $Q$.}
\end{itemize}
\TODO{any supporting sentences for the conjectures?}
These assumptions lead to the following expression for the transverse energy spectrum including MPI,
\begin{equation}
  \label{eq:MPIpre}
  \frac{d\sigma^{\text{pert+MPI}}}{dE_T} =  \frac{d\sigma^{\text{pert}}}{dE_T} \otimes f_{\text{MPI}}(E_T,\etacut)\;,
\end{equation}
where $f_{\text{MPI}}(E_T,\etacut)$ is the model function that needs to be fitted to the experiment. Similar approach was used in Refs.~\cite{Stewart:2014nna,Hoang:2017kmk,Kang:2018jwa} in order to incorporate for contribution from UE to jet substructure observables. We allow the model function to depend on $\etacut$ to properly incorporate  the change in phase-space for different experiments. The dependence on $\etacut$ can give us useful information regarding  the pseudo-rapidity distribution of MPI in hadronic collisions.

Note that the second conjecture can be relaxed allowing the model function to vary slowly with the hard scale. Then instead of Eq.(\ref{eq:MPIpre}), the transverse energy spectrum is given by
\begin{equation}
  \frac{d\sigma^{\text{pert+MPI}}}{dE_T} =  \int dQ \; \frac{d\sigma^{\text{pert}}}{dE_T dQ} \otimes f_{\text{MPI}}(E_T,Q,\etacut)\;,
\end{equation}
This approach might be more appropriate in studies over an extended range of $Q$. In this work we consider $Q\in(100,1000)$ GeV and in this region it is sufficient  to use the model of Eq.(\ref{eq:MPIpre}). For the parameterizations of the MPI model function we used the half normal distribution,
\begin{equation}
  \label{eq:modelF}
   f_{\text{MPI}}(E_T,\etacut) = \mathcal{N} \exp \lb -\lp\frac{E_T }{  \alpha(\etacut) \sqrt{\pi}} \rp^2\rb \Theta(E_T),
\end{equation}
where $\mathcal{N} = 2/(\alpha(\etacut)\,\pi )$ fixes the normalization of the model function to unity and $\alpha(\etacut)$ controls the first moment of the model function,
\begin{equation}
 \alpha(\etacut) = \langle E_T \rangle_{f} = \int_{0}^{\infty} d E_T\, E_T\, f_{\text{MPI}}(E_T,\etacut).
\end{equation}
If the conjectures above can be shown to be true up to power corrections, within the effective theory, then $\alpha(\etacut)$ can be written in terms of universal non-perturbative functions such as multiparton distribution functions. Here $\alpha(\etacut)$ can be fixed directly from experimental measurements using,
\begin{equation}
  \alpha(\etacut) =  \langle E_T \rangle_{\text{exp.}}-\langle E_T \rangle_{\text{pert.}}\,.
\end{equation}
Since no experimental data are available for this measurement (see Refs.~\cite{Aaboud:2017fwp,Aad:2016ria,Chatrchyan:2012tb,CMS:2016etb,Aad:2014hia} for relevant experimental studies) we use Monte Carlo simulation data. We find that in the region $1.5 <\etacut < 3.5$,  $\alpha(\etacut)$ can be well described by a linear fit,
\begin{equation}
  \label{eq:paramF}
\alpha(\etacut) = A \;\etacut\;,
\end{equation}
where $A$ is a parameter that describes the mean transverse energy deposited in the central region from MPI, and depends on the hadronic invariant mass, $\sqrt{s}$. Since in this work we are considering only $\sqrt{s}=13$ TeV we treat $A$ as a constant.   Fitting to the simulation data, we find $A=22.7$ GeV.
\begin{figure}[t!]
  \centerline{\includegraphics[width = \textwidth]{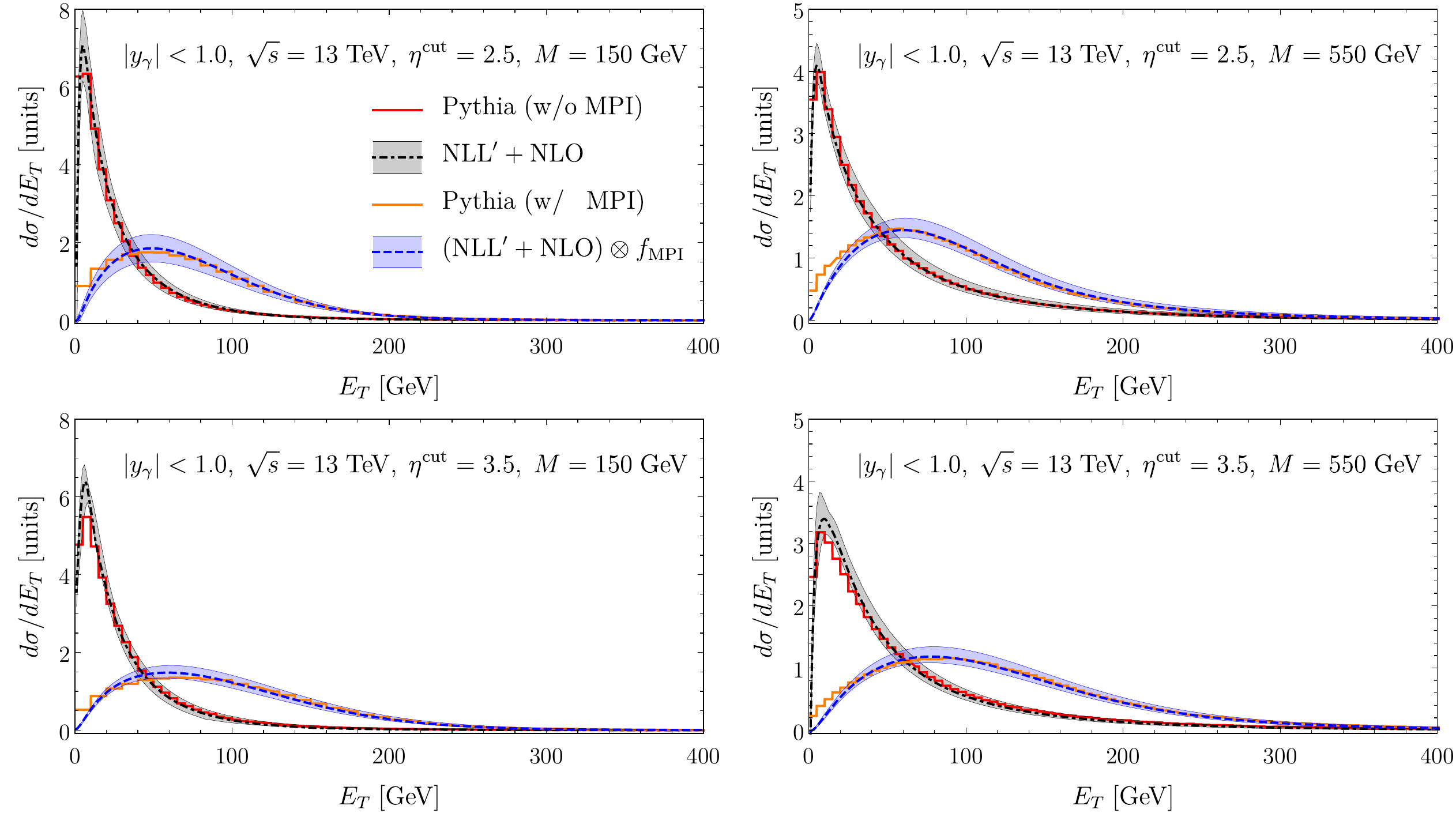}}
  \caption{The NLL-prime resumed distributions evaluated from Eq.(\ref{eq:matchingA}) and convolved with the half Gaussian model function (blue dashed) and the \textsc{Pythia} simulations (ISR+MPI) (orange solid). We also included the purely perturbative result (black dot-dashed) compared against the result from \textsc{Pythia} partonic distributions (MPI-only) (red solid). In each plot all curves are arbitrarily normalized to the same area.  }
\label{fig:dist}
\end{figure}

In Figure~\ref{fig:dist}, we illustrate the effect of MPI interactions to measurements of transverse energy within a pseudo-rapidity region, as described by \textsc{Pythia}. Once MPI effects are included, the transverse energy distribution differs significantly from the perturbation calculation. On the other hand, by includingthe contribution in Eq.(\ref{eq:MPIpre}) with the model in Eq.(\ref{eq:modelF}) we were able to accurately describe the simulation data.

We emphasize here that the aim of this section is to illustrate that a relatively simple model can describe the contribution of MPI for a large spectrum of the partonic invariant mass. More flexible models can achieve even better agreement, for example one can deviate from the linear fit in Eq.(\ref{eq:paramF}) allowing $A$ to depend on $\etacut$. Also we could deviate from he functional from of $f_{\text{MPI}}$ of  Eq.(\ref{eq:modelF}) (see also the work in Ref.~\cite{Papaefstathiou:2010bw} where the dependence of  $f_{\text{MPI}}$ on $s$ for fixed $\etacut =4.5$ is discussed).


\subsection{MPI-insensitive observables }
\label{sec:subtraction}
In this section we show how we  can use measurements of transverse energy to construct observables independent of MPI. Our proposal depends on the conjunctures above Eq.(\ref{eq:MPIpre}) thus the observables we propose can be used either in order to validate these conjunctures or for phenomenological studies, e.g., one can test the conjectures in Drell-Yan and use them in phenomenological studies of Higgs production.

We define the subtracted moments as follows
\begin{equation}
\label{eq:DeltaE}
\Delta \tra{E}^{(n)}(Q,Q_0) \equiv \langle\tra{E}^n(Q) \rangle -\langle \tra{E}^n(Q_0)\rangle
\end{equation}
where, $\langle\tra{E}^n(Q) \rangle$ is the  $n^{\text{th}}$ moment at the hard scale $Q$ and is defined by the following
\begin{equation}
\label{eq:ETn}
\langle\tra{E}^n(Q) \rangle =\int_0^\infty d\tra{E}\, \tra{E}^n\, \frac{d\sigma(E_T,Q)}{dE_T} \Big{/}\int_0^\infty d\tra{E} \frac{d\sigma(E_T,Q)}{dE_T}
\end{equation}
where $\sigma(E_T,Q)$ refers to  the differential cross section in $\tra{E}$ and $Q$. Assuming the MPI contribution can be modeled by a function $f_{\text{MPI}}(\tra{E})$ convoluted with the perturbative cross section in \eq{factII} we have,
\begin{equation}
\sigma(E_T,Q)= \int d\tra{E}'\,  f_{\text{MPI}}(\tra{E}-\tra{E}')\theta(\tra{E}-\tra{E}') \sigma^\text{pert} (\tra{E}',Q)
\end{equation} 
The numerator in \eq{ETn} can be written as
\begin{align}
&\int_0^\infty d\tra{E} \tra{E}^n  \int d\tra{E}'\, f_{\text{MPI}}(\tra{E}-\tra{E}') \theta(\tra{E}-\tra{E}') \sigma^\text{pert} (\tra{E}',Q)
\,,\nn\\
&= \int d\tra{E}'\, \sigma^\text{pert} (\tra{E}',Q) \int d\tra{E} (\tra{E}' +\omega)^n  f_{\text{MPI}}(\omega) \theta(\omega)
\,,\nn \\
&= \sum_{k=0}^{n} {}_nC_k \int d\tra{E}'\,(\tra{E}')^k \sigma^\text{pert} (\tra{E}',Q) \times \int d\omega \; \omega^{n-k}  f_{\text{MPI}}(\omega) \theta(\omega)
\end{align}
where in the second line  we  changed the integration variable to $\omega= \tra{E}-\tra{E}'$ and ${}_nC_k$ are the binomial coefficients. With this, \eq{ETn} can be written as
\begin{equation}
\label{eq:ETn-final}
\langle\tra{E}^n(Q) \rangle =\sum_{k=0}^{n} {}_nC_k\, \langle \tra{E}^k (Q) \rangle_\text{pert} \times \langle\tra{E}^{n-k} \rangle_\text{MPI}
\end{equation}
where the perturbative average $\langle \cdots \rangle_\text{pert}$ and MPI average $\langle \cdots \rangle_\text{MPI}$ are defined by \eq{ETn} with the replacement of the cross section by perturbative cross section $\sigma_\text{pert}(\tra{E},Q)$ and by the MPI model function $f_{\text{MPI}}(\tra{E})$ respectively. Applying \eq{ETn-final} in Eq.{\ref{eq:DeltaE}} we get
 \begin{align}
\Delta \tra{E}^{(n)} (Q,Q_0)
&=\sum_{k=1}^{n} {}_nC_k\, \left(  \langle \tra{E}^k (Q) \rangle_\text{pert} -\langle \tra{E}^k (Q_0) \rangle_\text{pert}  \right) \times \langle\tra{E}^{n-k} \rangle_{\text{MPI}}
\,,\nn\\
&=\sum_{k=1}^{n} {}_nC_k\, \Delta E_{T,\text{pert}}^{(k)}\, \times \langle\tra{E}^{n-k} \rangle_\text{MPI}
\end{align}
where $\Delta E_{T,\text{pert}}^{(k)}$ is defined in similar way to \eq{DeltaE} in terms of the perturbative cross section. By taking the difference at different values of $Q$, the first term at $k=0$, which is $\langle\tra{E}^{n} \rangle_\text{MPI}$, cancels. Thus for the first few values of $n$ we have,
\begin{align}
\label{eq:DeltaEfinal}
\Delta \tra{E}^{(1)} &= \Delta E_{T,\text{pert}}^{(1)}
\,,\nn\\
\Delta \tra{E}^{(2)} &= 2\Delta E_{T,\text{pert}}^{(2)}+\Delta E_{T,\text{pert}}^{(1)}\, \langle\tra{E} \rangle_\text{MPI}
\,,\nn\\
\Delta \tra{E}^{(3)} &=  3\Delta E_{T,\text{pert}}^{(3)}+3 \Delta E_{T,\text{pert}}^{(2)}\, \langle\tra{E} \rangle_\text{MPI}+ \Delta E_{T,\text{pert}}^{(1)}\, \langle\tra{E}^2 \rangle_\text{MPI}
\end{align}
Therefore, at $n=1$ the MPI contribution precisely cancels and the difference can be predicted by purely perturbative results. We refer to $\Delta \tra{E} \equiv \Delta \tra{E}^{(1)}$ as the subtracted transverse energy. The difference of higher moments includes MPI contributions that can be used to determine parameters of MPI models. Note that the results in \eq{DeltaEfinal} are obtained from the two conjectures above Eq.(\ref{eq:MPIpre}) but are independent of the model function, $f_{\text{MPI}}$.
\begin{figure}[t!]
  \centerline{\includegraphics[width = \textwidth]{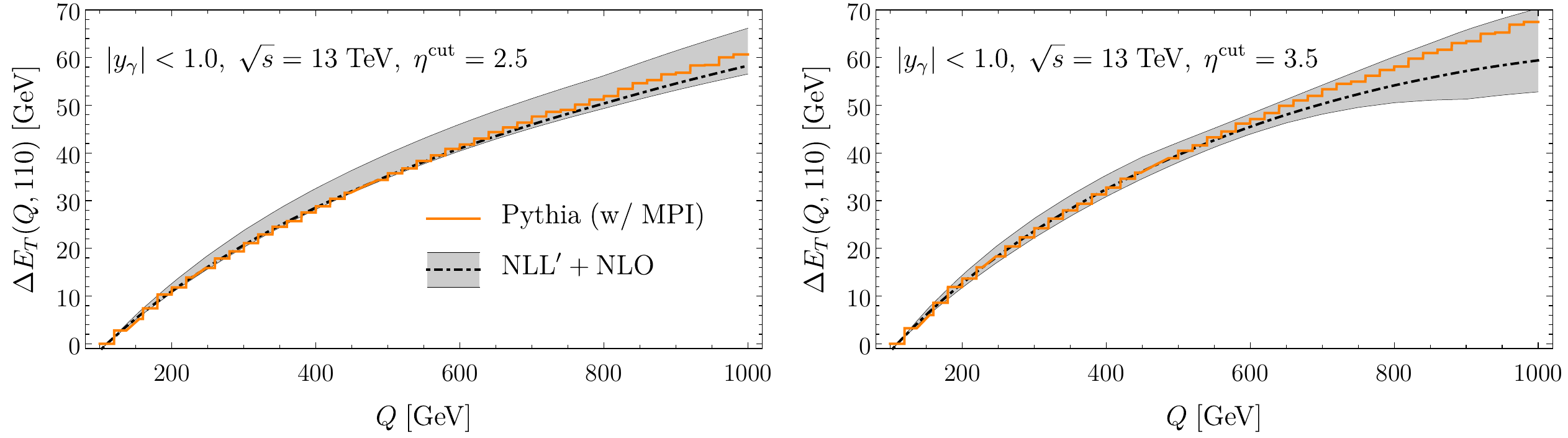}}
\caption{The observabel $\Delta E_T (Q,Q_0)$ for $Q_0=110$ GeV for two different values of the rapidity cutoff parameter $\etacut = 2.5$ (left) and $\etacut=3.5$ (right) }
\label{fig:DE}
\end{figure}

We demonstrate the cancelation of the MPI contribution in  $\Delta E_T$ using \textsc{Pythia} simulation in Figure~\ref{fig:DE}. We are comparing the observable $\Delta E_T (Q,110 \text{ GeV})$ evaluated with  the default MPI model of \textsc{Pythia} and the purely perturbative result. The uncertainty here is evaluated by using the maximum and minimum values of the error bands from the perturbative results.  It is clear that, within the uncertainty,  the observable $\Delta E_T$ is independent of the MPI contributions, as implemented in \textsc{Pythia}, for that range of the hard scale. In Figure~\ref{fig:DE} shows this for the cases $\etacut=2.5$ and $\etacut =3.5$.


\subsection{Generalization to other observables}
The approach of subtracting the mean at different hard scales to obtain MPI-insensitive measurements  can be implemented in other observables as well. This is true for additive observables for which the contributions from MPI can be  achieved through a convolution.  A characteristic example of this is the beam thrust~\cite{Stewart:2009yx,Stewart:2010pd,Berger:2010xi,Aad:2016ria}, $B$, defined as, \footnote{Here we use the definition provided in Ref.~\cite{Aad:2016ria}.}
\begin{equation}
  B=\sum_{i \in X} p_T^{(i)} \exp(-{\eta^{(i)}}).
\end{equation}
The corresponding  subtracted observable is
\begin{equation}
  \Delta B(Q,Q_0) \equiv \langle B(Q) \rangle -\langle B(Q_0) \rangle.
\end{equation}

\begin{figure}[h!]
  \centerline{\includegraphics[width = 0.5\textwidth]{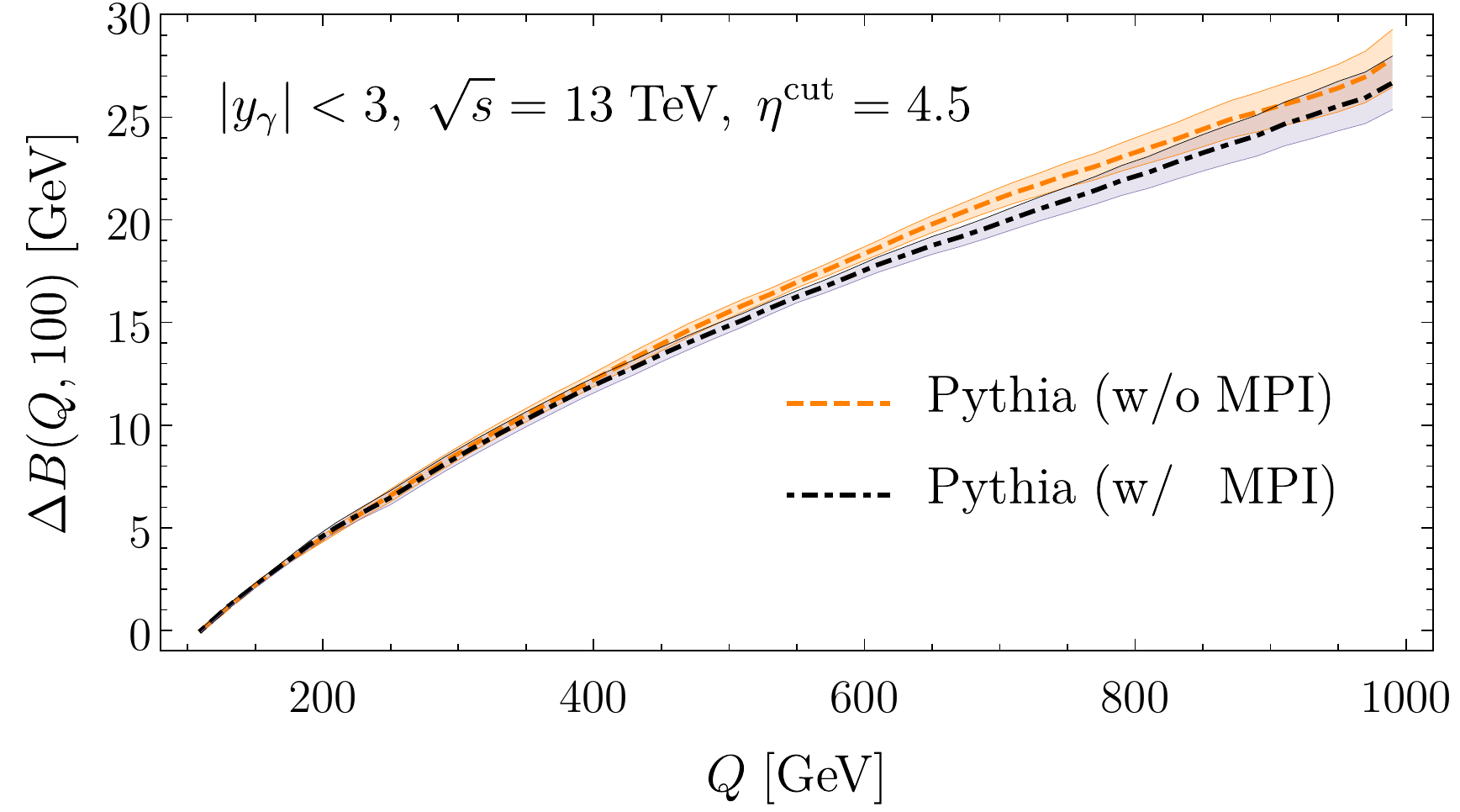}}
\caption{ The observable $\Delta B (Q,Q_0)$ (subtracted beam thrust) for  $Q_0=100$ GeV and $\etacut =4.5$. The uncertainty bands correspond to statistical uncertainty due to finite sample data.  }
\label{fig:DB}
\end{figure}

In Figure~\ref{fig:DB} we demonstrate that $ \Delta B(Q,Q_0)$ is also insensitive to the MPI contributions  for a large range of the virtual photon's invariant mass. An obvious advantage of using beam thrust is that the contribution of each particle is weighted by $\exp(-\eta)$ and thus contribution from particles in the forward region, close to the rapidity boundary, is exponentially suppressed. Therefore the measurement is insensitive to the rapidity cutoff~\footnote{This assumes that the cutoff does not enter the central region, i.e., $\etacut \gtrsim 1.5$}. On the other hand this means that we cannot use beam thrust as our observable if we aim to  study the pseudo-rapidity dependence of MPI through the model function $f_{\text{MPI}}$.

\section{Conclusion}
\label{sec:conclusions}
In this paper we demonstrated  an effective field theory approach for calculating the  transverse energy spectrum when a rapidity cutoff is imposed. In contrast to Ref.~\cite{Hornig:2017pud}, in this work we use dynamical (profile) scales in order to interpolate between our result from the small transverse energy region (region II) to large values of transverse energy (region I). Finally we used a subtraction scheme to match onto the fixed order result of the full theory (QCD) at $E_T \sim Q$.  Although the cross section in the far tail region where $E_T \sim Q$ is highly suppressed compared to regions I and II, it is important to know the spectrum in all ranges of $E_T$ since the far tail region gives significant contributions when we calculate moments of the transverse energy.

As an example we choose to study the process $pp \to \gamma(\to \ell^+ \ell^-) +X$ away from the $Z$-pole region. Comparing our results with \textsc{Pythia} (ISR only) simulations we find excellent agreement. We then proceed to introduce a prescription to include the effect of multiparton  interactions (MPI). The prescription we propose, which is based on the conjectures above Eq.(\ref{eq:MPIpre}), is to simply convolve the perturbative spectrum  with a model function. The model function should not depend on the hard process but could have small variations with the change of the hard scale in the process. Comparing our result with \textsc{Pythia} (ISR+MPI) we find that for the range 100-1000 GeV of the photon's invariant mass the MPI model function has little or no hard scale dependence.

Assuming  independence of the model function on the hard scale of the problem we introduce an observable, which is independent of MPI effects. This observable we are considering is the subtracted first moment of the transverse energy at two different scales. We compare our purely perturbative calculations of this observable against \textsc{Pythia} (ISR+MPI) and we show that the two agree very well.

Although in this paper we considering only transverse energy measurements, in the last section we discuss generalizations of this MPI insensitive observable to other event shapes, such as beam thrust and transverse thrust. An advantage of using transverse energy as a probe to the MPI effects is that we have strong sensitivity to the rapidity cutoff parameter, $\etacut$. Measurements at different values of $\etacut$ can give an insight into the pseudo-rapidity dependence of MPI.

The independence of the MPI to the hard process using Monte Carlo  simulations was also investigated in Ref.~\cite{Papaefstathiou:2010bw} in  Higgs, $Z$, and $W^{\pm}$ production processes implemented in {\ttfamily Herwig++}~\cite{Bahr:2008pv}. A future application of our work is to evaluate the Higgs or $Z$/$W^{\pm}$ transverse energy spectrum, using the prescription and model function for MPI we propose in this paper, and compare with results from Monte Carlo simulations.

\acknowledgments
YM would like to thank Duff Neill and Varun Vaidya for useful discussions. YM is supported by the DOE Office of Science under Contract DE-AC52-06NA25396 and the Early Career Research Program (C. Lee, PI), and through the LANL/LDRD Program.
TM is supported in part by the Director, Office of Science, Office of Nuclear Physics, of the U.S. Department of Energy under grant numbers DE-FG02-05ER41368.
DK would like to thank the Nuclear Theory group at LANL for hospitality during portions of this work.


\appendix

\section{The beam function matching for region I}

\label{ap:A}
We evaluate the bare global beam function $\mathcal{O}(\alpha_s)$ terms using Eq.(25) and (26) from Ref.~\cite{Ritzmann:2014mka},
\begin{equation}
  \label{eq:A.1}
  \mathcal{B}^{\text{ G (1,b)}}_{q/q} (x,E_{T};\mu,\nu) = \frac{\alpha_s C_F}{\pi} \frac{\exp(\epsilon \gamma_E)}{\Gamma(1-\epsilon)} \lp \frac{\nu}{p^{-}} \rp ^{\eta} \frac{1}{\mu} \lp \frac{\mu}{E_T}\rp ^{1+2\epsilon} \lb \frac{1+x^2} {(1-x)^{1+\eta}} - \epsilon (1-x)^{1-\eta}  \rb.
\end{equation}
Expanding first in $\eta$ and then in $\epsilon$ we have,
\begin{multline}
  \mathcal{B}^{\text{ G (1,b)}}_{q/q} (x,E_{T};\mu,\nu) = \frac{\alpha_s C_F}{\pi} \lbc -\frac{1}{\eta} \lb -\frac{1}{\epsilon} \delta(E_T) +2 \mathcal{L}_0 (E_T,\mu)  \rb \delta(1-x) \\+\frac{1}{\epsilon} \lb \ln \lp \frac{\nu}{p^-} \rp \delta(1-x) -\frac{1}{2} \bar{P}_{q/q} (x)   \rb  \delta(E_T) +\mathcal{L}_0 (E_T, \mu) \lb \bar{P}_{q/q} (x) -2\ln \lp \frac{\nu}{p^-} \rp  \delta(1-x)     \rb  \\
  + c_{q/q} (x) \delta(E_T) \rbc,
\end{multline}
where
\begin{equation}
  c_{q/q}(x) = \frac{1-x}{2},
\end{equation}
and
\begin{equation}
\bar{P}_{q/q}(x) = \frac{1+x^2}{(1-x)_+} - \frac{3}{2} \delta(1-x) =P_{q/q}(x) -\bar{\gamma}_q \;\delta(1-x),
\end{equation}
where $P_{i/j}(x)$ are the QCD splitting kernels~\cite{Altarelli:1977zs, PhysRevD.9.980} and  $\bar{\gamma}_q =3/2$. The divergent $P_{q/q}(x)/\epsilon$  term cancels during the matching with the collinear parton distribution functions (PDFs) and therefore should not be included in the renormalization kernel. 
Thus in the $\overline{\text{MS}}$  scheme this yields,
\begin{multline}
  \label{eq:Iqq1}
  \mathcal{I}^{\text{ G (NLO)}}_{q/q} (x,E_T,p^-) = \delta(E_T) \delta(1-x) +\frac{\alpha_s C_F}{\pi} \lbc \mathcal{L}_0 (E_T, \mu) \lb \bar{P}_{q/q} (x) -2\ln \lp \frac{\nu}{p^-} \rp   \delta(1-x)     \rb \\ + c_{q/q} (x) \delta(E_T) \rbc,
\end{multline}
and the corresponding renormalization function
\begin{equation}
  Z_{q/q}^{\text{B,G (1)}} (E_T,p^-)  = \delta(E_T) +\frac{\alpha_s C_F}{\pi} \lbc -\frac{1}{\eta} \lb 2 \mathcal{L}_0 (E_T,\mu) -\frac{1}{\epsilon} \delta(E_T)  \rb +\frac{1}{\epsilon} \lb \ln \lp \frac{\nu}{p^-} \rp +\frac{1}{2} \bar{\gamma}_q    \rb  \delta(E_T) \rbc,
  \end{equation}
where  we omitted  the rapidity and virtuality scale the arguments for simplicity of notation. To evaluate the off-diagonal element $\mathcal{I}_{q/g}^{\text{ G}}$ we start with the corresponding partonic beam function given in Eq.(26) of Ref.~\cite{Ritzmann:2014mka},
\begin{equation}
  \label{eq:A.7}
  \mathcal{B}_{q/g}^{\text{ G (1,b)}}(x,E_T,p^-;\mu,\nu)= \frac{\alpha_s T_F}{\pi} \frac{e^{\epsilon \gamma_E} }{\Gamma(2-\epsilon)} \lp \frac{\nu}{p^{-}} \rp ^{\eta} \frac{1}{\mu} \lp \frac{\mu}{E_T}\rp ^{1+2\epsilon} (1-x)^{-\eta} \lb P_{q/g}(x) -\epsilon  \rb.
\end{equation}
where
\be
P_{q/g}(x)= x^2+(1-x)^2 
\ee
Expanding first in $\eta$ and then in $\epsilon$ we have,
\begin{equation}
  \mathcal{B}_{q/g}^{\text{ G (1,b)}}(x,E_T,p^-;\mu,\nu)= \frac{\alpha_s T_F}{\pi} \lbc - \frac{1}{2\epsilon} P_{q/g} (x) +\mathcal{L}_0 (E_T,\mu) P_{q/g}(x) +c_{q/g}(x) \delta(E_T)  \rbc,
\end{equation}
where
\begin{equation}
  c_{q/g}(x) = x(1-x).
\end{equation}
Except the term $P_{q/g}(x) /\epsilon$ which cancels during the matching there is no other divergent term. Thus the contribution from the off-diagonal element at this order does not contribute to the renormalization function or the corresponding anomalous dimension. Thus the matching coefficient is,
\begin{equation}
   \label{eq:Iqg1}
  \mathcal{I}_{q/g}^{\text{ G (NLO)}} = \frac{\alpha_s T_F}{\pi} \lbc  \mathcal{L}_0 (E_T,\mu) P_{q/g}(x) +c_{q/g}(x) \delta(E_T)  \rbc.
\end{equation}

 The rapidity and virtuality anomalous dimensions, $\gamma_{\nu}$ and $\gamma_{\mu}$ respectively are evaluated using the following,
\begin{align}
  \gamma_{\nu,q}^{\text{B,G}}(E_T;\mu)& =-( Z^{\text{B,G}} _{q/q} )^{-1} \otimes \frac{d}{d\ln(\nu)}  Z^{\text{B,G}} _{q/q} ,  \\
  \gamma_{\mu,q}^{\text{B,G}}(p^-;\mu,\nu) \delta(E_{T})& =-( Z^{\text{B,G}} _{q/q} )^{-1} \otimes \frac{d}{d\ln(\mu)}  Z^{\text{B,G}} _{q/q} ,
  \end{align}
and thus we have
\begin{align}
  \gamma_{\nu,q}^{\text{B,G}}(E_T;\mu)& = - \frac{2 \alpha_s(\mu) C_F}{\pi} \mathcal{L}_0(E_T,\mu),   \\
  \gamma_{\mu,q}^{\text{B,G}}(p^-;\mu,\nu) & = \frac{\alpha_s(\mu) C_F}{\pi} \lb 2 \ln \lp \frac{\nu}{p^-} \rp + \bar{\gamma}_q \rb. 
\end{align}


\section{Merging fixed order}
\label{ap:B}

Here we perform an explicit  calculation to show that the product $S_{n}\otimes \cB^{\text{ II}}_{a/P}$ reduces to $\cB^{\text{ G}}_{a/P}$ in the large transverse energy limit, $E_T \gg Qr$. To this end, we work with the partonic-level functions, $\cB^{\text{ G}}_{q/i}$ in \eq{factI} and the combination of soft-collinear and beam function $S_{n}\otimes \cB^{\text{ II}}_{q/i}$ in Eq.(\ref{eq:factII}) where $i=q,g$. At $\mathcal{O}(\alpha_s^1)$ we have
\begin{multline}
  S_n(E_T,r;\mu,\nu) \otimes\cB^{\text{ II}}_{q/i}(x,E_T,r;\mu) = \delta_{qi}\delta(E_T) \delta(1-x) \\[3pt]
  + S^{(1)}_n(E_T,r;\mu,\nu) \delta_{qi} \delta(1-x) + \mathcal{I}_{q/i}^{\text{ II} \;(1)}(x,E_T,r;\mu) + \mathcal{O}(\alpha_s^2),
\end{multline}
and
\begin{equation}
  \cB^{\text{ G}}_{q/P}(x,E_T;\mu,\nu) = \delta_{qi}\delta(E_T) \delta(1-x) + \mathcal{I}_{q/i}^{\text{ G} \;(1)}(x,E_T;\mu,\nu) + \mathcal{O}(\alpha_s^2),
\end{equation}
therefore we need to show
\begin{equation}
  \label{eq:merge-pc}
  \lb S^{(1)}_n(E_T,r;\mu,\nu) \delta_{qi} \delta(1-x) + \mathcal{I}_{q/i}^{\text{ II} \;(1)}(x,E_T,r;\mu) \rb  \xrightarrow{E_T \gg Qr} \;  \mathcal{I}_{q/i}^{\text{ G} \;(1)}(x,E_T;\mu,\nu) + \text{p.c.}
  \end{equation}
This task becomes much easier if we work with cumulant bare functions defined 
\begin{equation}
\label{Fcum}
F(\veb{p}) = \int_0^{\veb{p}} F(\tra{E})
\end{equation}
where the cumulant of the one-loop beam function in \eq{factI} is given by integrating Eq.(\ref{eq:A.1}),
  \begin{equation}
  \label{eq:BqqI}
 \mathcal{I}_{q/q}^{\text{ G} \;(1)}(x, \veb{p}) = \frac{\as C_F}{\pi}\frac{e^{\eps \gamma_E}}{\Gamma(1-\eps)} \lp \frac{\nu}{p^-} \rp ^\eta 
  \lb \frac{1+x^2}{(1-x)^{1+\eta}}-\eps (1-x)^{1-\eta}\rb
  \frac{-1}{2\eps}\lp\frac{\mu}{\veb{p}}\rp^{2\eps}.
\end{equation}  
The one-loop cumulant soft-collinear function is given integrating Eq.(2.12) of Ref.\cite{Hornig:2017pud}
  \begin{equation}
  \label{eq:Sn}
  S^{(1) }_{n}(\veb{p},r) =- \frac{\as C_F}{2\pi}\frac{4e^{\eps \gamma_E}}{\Gamma(1-\eps)}
  \frac{1}{\eta} \lp\frac{\nu r}{\mu} \rp^\eta 
  \frac{-1}{2\eps+\eta}\lp\frac{\mu}{\veb{p}}\rp^{2\eps+\eta} 
\,,\end{equation}  
 and the one-loop cumulant beam function consists of in-jet, out-of-jet, the zero-bin contributions, given by integrating Eqs.(B.1), (B.10), and (B.13) of Ref.~\cite{Hornig:2017pud} respectively,     
   \begin{equation}
    \label{eq:BqqII}
 \mathcal{I}_{q/q}^{\text{ II} \;(1)}(x,\veb{p},r) = \mathcal{I}_{q/q}^{(1)}(x,r)+(  \Delta B^\text{out}_{q/q} (x,\veb{p},r) -  \Delta B^\text{zero}_{q/q} (x,\veb{p},r) )\,,
   \end{equation}
   where
  \begin{align}
\mathcal{I}_{q/q}^{(1)}(x,r) &= - \frac{\as \CF}{2\pi}\frac{e^{\eps \gamma_E} x^{2\eps}}{\eps\Gamma(1-\eps)} \lp \frac{\mu}{p^- r}\rp^{2\eps}
   \lb \frac{1+x^2}{(1-x)^{1+2\eps}}-\eps (1-x)^{1-2\eps}\rb
  \,,\nn\\
 \Delta B^\text{out}_{q/q} (x,\veb{p},r)&= \frac{\as \CF}{2\pi}\frac{e^{\eps \gamma_E} }{\eps\Gamma(1-\eps)} \lp \frac{\nu}{p^-} \rp^\eta 
    \lb \frac{1+x^2}{(1-x)^{1+\eta}}-\eps (1-x)^{1-\eta}\rb
 \nn\\ & \quad \times   \theta(x-x_0) \lb \lp \frac{\mu}{p^- r}\frac{x}{1-x}\rp^{2\eps}-\lp \frac{\mu}{ \veb{p}} \rp ^{2\eps}  \rb , \nn\\[5pt]
 \Delta B^\text{zero}_{q/q} (x,\veb{p},r) &=\delta(1-x)\, S^{(1) }_{n}
   \,,
\end{align}
where $x_0= [1+\veb{p}/(p^- r)]^{-1} $.  Note that  the zero-bin contribution on the last line is precisely cancelled against  the soft-collinear function in \eq{Sn}.
\begin{align}
\label{eq:SBfinal}
   S^{(1) }_{n}(\veb{p},r) \delta(1-x) &+ \mathcal{I}_{q/q}^{\text{ II} \;(1)}(x,\veb{p},r)\nn \\ &=\mathcal{I}_{q/q}^{(1)}(x,r) +  \Delta B^\text{out}_{q/q} (x,\veb{p},r)
 \nn\\
 &= \mathcal{I}_{q/q}^{\text{ G} \;(1)}(x, \veb{p}) +\Delta B^\text{out}_{q/q} (x,\veb{p},r) \Big{\vert}_{\theta(x-x_0)\to \theta(x-x_0)-\theta(x)}\;.
 \end{align}
In the last step, we use the identify:
\begin{equation}
\Delta B^\text{out}_{q/q} (x,\veb{p},r) = \theta(x-x_0) \lb -\mathcal{I}_{q/q}^{(1)}(x,r)+ \mathcal{I}_{q/q}^{\text{ G} \;(1)}(x, \veb{p}) \rb+ \mathcal{O}(\eta)\;.
\end{equation}
Note that the second term on last line is simply power corrections in the limit $\veb{p}\ll p^- r$:
\begin{equation}
\label{eq:thetaxx0}
\theta(x-x_0)-\theta(x) \to -\lp \frac{p^- r}{\veb{p} }\rp \delta (x)+O\lb \lp\frac{p^- r}{\veb{p}}\rp^2 \rb\;.
\end{equation}
This shows the diagonal element (i.e., $i=q$) of Eq.({\ref{eq:merge-pc}) up to power corrections. 

The gluon channel is similar but simpler.
The gluon channel contribution in \eq{factI} can be found integrating Eq.(\ref{eq:A.7}) and setting $\eta \to 0$
\begin{equation}
\label{eq:BqgI}
\mathcal{I}_{q/g}^{\text{ G}\;(1)}(x,\veb{p})=-\frac{\as \TF}{2\pi} \frac{e^{\eps\gamma_E}}{\epsilon\;\Gamma (2-\eps)} \lb P_{qg}(x)-\eps\rb \lp \frac{\mu}{\veb{p}}\rp^{2\eps}\;.
\end{equation}
There is no zero-bin contribution in the gluon channel thus using Eqs.(B.4) and (B.18) of Ref.~\cite{Hornig:2017pud}
\begin{equation}
\label{eq:BqgII}
\mathcal{I}_{q/g}^{\text{ II} \;(1)}(x,\veb{p},r) = \mathcal{I}_{q/g}^{(1)}(x,r)+  \Delta B^\text{out}_{q/g} (x,\veb{p},r) \,,
\end{equation}
where
\begin{align}
  \mathcal{I}_{q/g}^{(1)}(x,r) &= -\frac{\as \TF}{2\pi} \lp \frac{1}{\eps}+1\rp \lp\frac{x}{1-x}\rp^{2\eps} \lp \frac{\mu}{p^- r}\rp^{2\eps} \left[ P_{qg}(x)-\eps\right]
  \,,\nn\\
   \Delta B^\text{out}_{q/g} (x,\veb{p},r) &= \frac{\as \TF}{\pi}\frac{1}{2\eps(1-\eps)} \left[ P_{qg}(x)-\eps\right] 
  \theta(x-x_0) \lb \lp \frac{\mu}{p^- r} \frac{x }{1-x}\rp^{2\eps}-\lp \frac{\mu}{ \veb{p} } \rp^{2\eps}  \rb \;.
\end{align}  
By comparing \eq{BqgI} to \eq{BqgII}, we find
\begin{equation}
  \Delta B^\text{out}_{q/g} (x,\veb{p},r) = \theta(x-x_0)\lb  \mathcal{I}_{q/g}^{\text{ G}\;(1)}(x,\veb{p})   -  \mathcal{I}_{q/g}^{(1)}(x,r) \rb + \mathcal{O}(\eps)\;,
\end{equation}
and we have
\begin{equation}
\label{eq:Bqgfinal}
\mathcal{I}_{q/g}^{\text{ II} \;(1)}(x,\veb{p},r) = \mathcal{I}_{q/g}^{\text{ G}\;(1)}(x,\veb{p}) + \Delta B^\text{out}_{q/g} (x,\veb{p},r) \Big{\vert}_{\theta(x-x_0)\to \theta(x-x_0)-\theta(x)} \;.
\end{equation}
As in \eq{SBfinal}, the second term on RHS is the power correction.


\section{Evolution and resummation}
\label{ap:C}
In this appendix we give the details for the solutions of the renormalization group and rapidity renormalization group equations. This section is divided into two subsections. In Section~\ref{ap:C1} we discuss the virtuality renormalization group equations and the solutions of those equations and in Section~\ref{ap:C2} the rapidity renormalization group evolution is described. All elements of factorization (hard, soft, soft-collinear, and beam) satisfy renormalization group equations, in contrast, only transverse energy dependent quantities  have rapidity RGE.


\subsection{Renormalization group evolution }
\label{ap:C1}

The RGEs we encounter in this work belong to the same category of what was referred to in Ref.~\cite{ Hornig:2017pud} as unmeasured evolution equations. In this paper we do not discuss the evolution of measured quantities and therefore such a distinction is redundant. Also for the processes we are considering the hard and soft function have trivial color structure and therefore we do not address the complications that appear when one considers multi-jet processes in hadronic collisions. The RGEs we consider have the following form,
\begin{equation}
  \label{eq:unmeasRG}
   \frac{d}{d \ln\mu}F(\mu)= \gamma_{\mu}^{F} (\mu, \alpha_s) F(\mu)= \lb \Gamma^{F}_{\mu}[\alpha_S] \ln \lp \frac{\mu^2}{m^2_F} \rp + \Delta \gamma^F_{\mu}[\alpha_S]\rb F(\mu),
\end{equation}
where $\gamma_{\mu}^{F}$ is the virtuality anomalous dimension. We refer to the first term in the square brackets as the cusp part since $\Gamma^{F}_{\mu}[\alpha_s]$ is proportional to the cusp anomalous dimension, and the second term, $\Delta \gamma^F_{\mu}[\alpha_S]$, as the non-cusp part. Both the cusp and the non-cusp terms have an expansion in the strong coupling. For the cusp term we have, 
\begin{equation}
  \label{eq:G}
\Gamma_F[\alpha_s] =  (\Gamma_F^0/\Gamma_{\text{cusp}}^0) \Gamma_{\text{cusp}} = (\Gamma_F^0/\Gamma_{\text{cusp}}^0) \sum_{n=0}^{\infty} \left(\frac{\alpha_s}{4 \pi} \right)^{1+n} \Gamma_{\text{cusp}}^n,
\end{equation}
and similarly the non-cusp part is given by,
\begin{equation}
   \label{eq:g}
\gamma_F[\alpha_s] =  \sum_{n=0}^{\infty} \left(\frac{\alpha_s}{4 \pi} \right)^{1+n} \gamma_{F}^n,
\end{equation}
The solution to the RGE in Eq.(\ref{eq:unmeasRG}) is
\begin{align}
  \label{eq:U}
F(\mu)&= \mathcal{U}_F(\mu,\mu_0) F(\mu_0) \, , &\mathcal{U}_F(\mu,\mu_0)=\exp \left( K_F (\mu, \mu_0) \right) \left( \frac{\mu_0}{m_F} \right) ^{\omega_F(\mu, \mu_0)},
\end{align}
with
\begin{align}
  \label{eq:Kt}
K_F(\mu, \mu_0) &= 2 \int_{\alpha (\mu_0)}^{\alpha(\mu)} \frac{d \alpha}{\beta(\alpha)} \Gamma_F (\alpha) \int_{\alpha(\mu_0)}^{\alpha}
\frac{d \alpha'}{\beta(\alpha')} +\int_{\alpha (\mu_0)}^{\alpha(\mu)} \frac{d \alpha}{\beta(\alpha)} \gamma_F (\alpha) ,\\
\label{eq:wt}
\omega_F(\mu, \mu_0) &= 2 \int_{\alpha (\mu_0)}^{\alpha(\mu)} \frac{d \alpha}{\beta(\alpha)} \Gamma_F (\alpha).
\end{align}
Since in this work we are interested only in the NLL and NLL' result we may keep only the first two terms in the perturbative expansion of the cusp part (i.e., $\Gamma_0^F$, $\Gamma^{0}_{\text{cusp}}$, and $\Gamma^{1}_{\text{cusp}}$) and only the first term form the non-cusp part ($\gamma_{F}^{0}$). Performing this expansion we get,
\begin{align}
    \label{eq:K}
K_F(\mu, \mu_0) &=-\frac{\gamma_F^0}{2 \beta_0} \ln r -\frac{2 \pi \Gamma_F^0}{(\beta_0)^2} \Big{\lbrack} \frac{r-1+r\ln r}{\alpha_s(\mu)}
+ \left( \frac{\Gamma^1_{\text{cusp}}}{\Gamma^0_{\text{cusp}}}-\frac{\beta_1}{\beta_0} \right) \frac{1-r+\ln r}{4 \pi}+\frac{\beta_1}{8 \pi \beta_0}
\ln^2 r  \Big{\rbrack}, \\
\label{eq:w}
\omega_F(\mu, \mu_0) &= - \frac{\Gamma_F^0}{ \beta_0} \Big{\lbrack} \ln r + \left( \frac{\Gamma^1_{\text{cusp}}}{\Gamma^0_{\text{cusp}}} -
\frac{\beta_1}{\beta_0}  \right) \frac{\alpha_s (\mu_0)}{4 \pi}(r-1)\Big{\rbrack},
\end{align}
where $r=\alpha(\mu)/\alpha(\mu_0)$ and $\beta_n$ are the coefficients of the QCD $\beta$-function,
\begin{equation}
\beta(\alpha_s) = \mu \frac{d \alpha_s}{d \mu}= -2 \alpha_s \sum_{n=0}^{\infty} \left( \frac{\alpha_s}{4 \pi} \right)^{1+n} \beta_n \; .
\end{equation}
Table~\ref{tb:evolution} the expressions for all ingredients necessary to perform the evolution of any function that appears in the factorization theorems we considered in this paper are given in.
\begin{table}[t!]
  \renewcommand{\arraystretch}{1.3}
  \begin{center}
\begin{tabular}{|c|c|c|c|c|c|}
\hline
Function  & $H_{a\bar{a}}$ & $S_s^{a\bar{a}}$ & $S_{n}^{a}$  & $\mathcal{B}_{a/P}^{\text{G}}$ & $\mathcal{B}_{a/P}^{\text{II}}$\\
\hline \hline
$\Gamma_{F}^{0}$ & $-4(C_a+C_{\bar{a}})$ & $ 4(C_a+C_{\bar{a}})$ & $ -4 C_a$ & 0 & $ 4 C_a$ \\\hline
$\gamma_{F}^{0}$ & $-4 \bar{\gamma}_a (C_a+C_{\bar{a}})$ & 0 & 0 & $4 \pi \gamma_{\mu,a}^{\text{B, G}}/\alpha_s$ & $ 4 C_a \bar{\gamma}_a$ \\\hline
$m_{F}$ & $Q$ & $\nu_{ss}$ & $\nu_{sc} r$ & n.a. & $  p^- r $ \\\hline
$\xi_{F}$ & n.a. & $ 4(C_a+C_{\bar{a}})$ & $ -4 C_a$ & $ -4 C_a$ & n.a.  \\\hline
$\Delta\gamma_{\nu}^{F}$ & n.a. & $ \mathcal{O}(\alpha_s)$ & $ \mathcal{O}(\alpha_s)$ & $ \mathcal{O}(\alpha_s)$ & n.a.  \\\hline
\end{tabular}
 \caption{Evolution table: $\bar{\gamma}_q = 3/2$ and $\gamma_{\mu,a}^{\text{B, G}} = (\alpha_s (\mu) C_F /\pi) \lp 2\ln (\nu/p^-)  + \bar{\gamma}_q \rp$.}
  \label{tb:evolution}
\end{center}
\end{table}


\subsection{Rapidity renormalization group evolution }
\label{ap:C2}

In this section we summarize the solution for the rapidity renormalization group equations for the global soft, $S_s$, soft-collinear, $S_n$, and global beam, $\mathcal{B}_{a/P}^{\text{G}}$ functions. Even though the unmeasured beam function of region II has transverse energy dependence, it does not have rapidity divergences and thus does not acquire rapidity scale dependence. The RRG equation for transverse energy measurements of the function $F(E_T) \in \{ S_s, S_n,\mathcal{B}^{\text{G}}_{a/P} \}$ takes the following form,
\begin{equation}
  \frac{d}{d\ln\nu} F(E_T,\mu,\nu) = \int dE_T'\; \gamma_{\nu}^{F} (E_T - E_T',\mu) F( E_T',\mu,\nu)  \equiv  \gamma_{\nu}^{F} \otimes  F( E_T,\mu,\nu),
  \end{equation}
where
\begin{equation}
  \label{eq:rg}
  \gamma_{\nu}^{F} (E_T,\mu) = 2\Gamma_{\nu}^F [\alpha_s] \mathcal{L}_{0} (E_T,\mu) +\Delta\gamma_{\nu}^F [\alpha_s].
\end{equation}
The solution of this equations is
\begin{equation}
  F(E_T,\mu,\nu) = \int dE_T \; \mathcal{V}_F(E_T -E_T',\mu,\nu,\nu_0) F(E_T',\mu,\nu_0),
\end{equation}
where
\begin{equation}
\label{eq:RRGEsoln}
  \mathcal{V}_F(E,\mu,\nu,\nu_0) =  \frac{e^{\kappa_F(\mu,\nu,\nu_0)}(e^{\gamma_E} \mu)^{-\eta_F(\mu,\nu,\nu_0)}}{\Gamma(\eta_F(\mu,\nu,\nu_0))}  \lb \frac{1}{E^{1-\eta_F(\mu,\nu,\nu_0)}} \rb_+,
\end{equation}
and 
\begin{align}
  \eta_F(\mu,\nu,\nu_0) &= 2 \Gamma_{\nu}^F[\alpha] \ln \left( \frac{\nu}{\nu_0}  \right),  &  \kappa_F(\mu,\nu,\nu_0) &=  \Delta\gamma_{\nu}^F[\alpha] \ln \left( \frac{\nu}{\nu_0} \right),
\end{align}
where $\nu_0$ is the characteristic scale (for each function) from which we start the evolution. This scale is chosen such that rapidity logarithms are minimized. For the global-soft, soft-collinear, and beam functions these scales are given in Table~\ref{tb:evolution}. The first term in the rapidity anomalous dimension in Eq.(\ref{eq:rg}) is proportional to the cusp anomalous dimension and the proportionality constant we denote with $\xi_F$,  ($\Gamma_{\nu}^{F} = \xi_F \Gamma_{\text{cusp}}$). We define the the plus-distribution in Eq.~(\ref{eq:RRGEsoln}) through its inverse Laplace transform,
\bea
  \lb \frac{1}{E^{1-\alpha}} \rb_+ =  {\cal L}^{-1}\lb s^\alpha \, \Gamma[-\alpha]\rb \nn \; .
  \eea
  
\section{Profile functions and scale variation}
\label{ap:profiles}

\begin{figure}[t!]
  \centerline{\includegraphics[width = \textwidth]{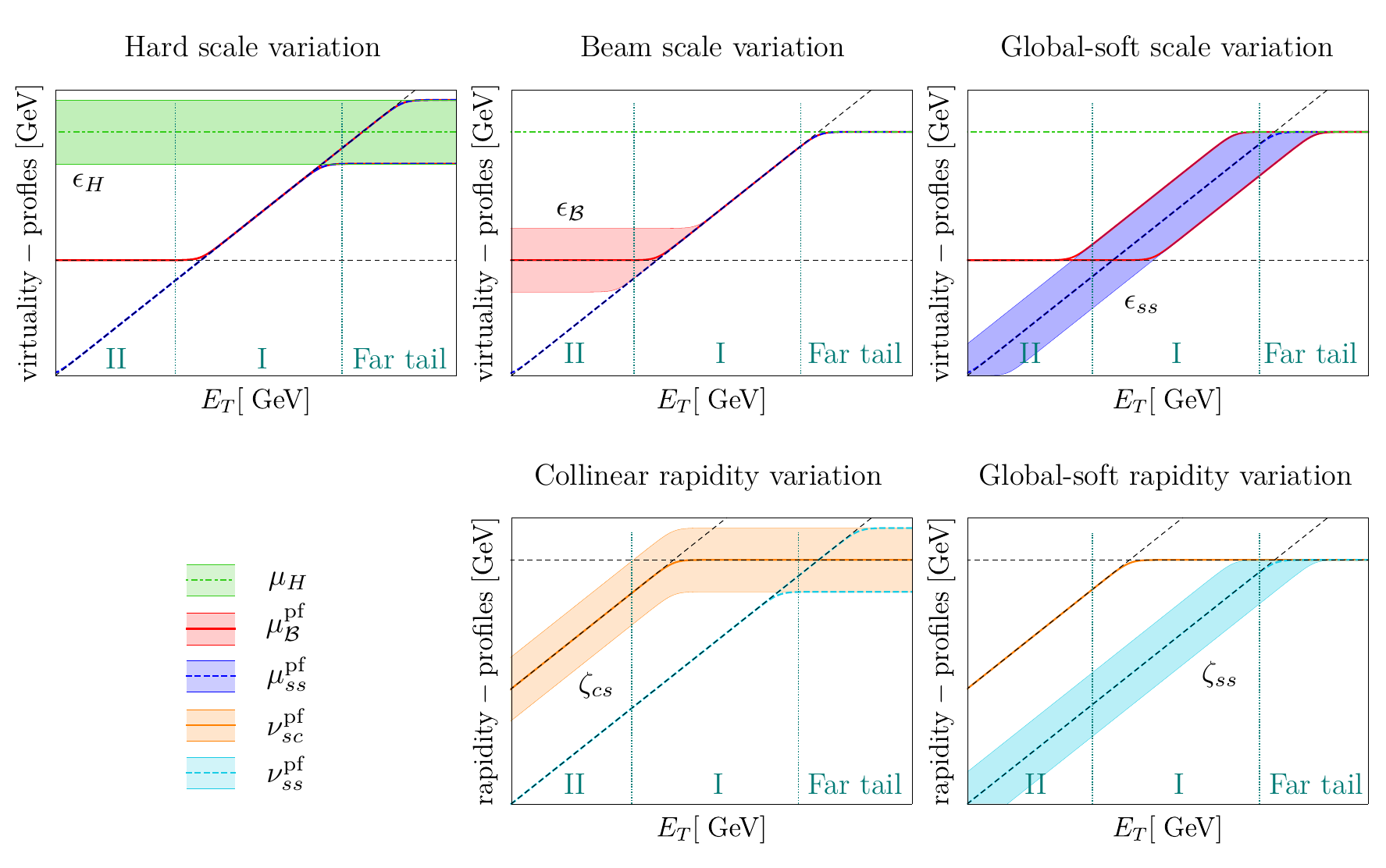}}
\caption{Illustration of the profile scales and how each of the four profile functions changes with the five scale variations. For improving readability we use log-log plots.}
\label{fig:scales}
\end{figure}

In this section we give the details on the choice of profile functions and how scale variation is performed. There are four profile functions. They assist either for the transition from region I to region II or for turning off evolution in the far tail (in order to match onto the fixed order result). The asymptotic behavior in all regions for all profile functions is collected in Eqs.(\ref{eq:profiles}) and (\ref{eq:profiles2}). The relations in Eq.(\ref{eq:profiles}) need to be always satisfied in order to reproduce the correct result when transitioning from region II to region I. That is, for example, when we vary $\mu_{ss}$ (in order to explore uncertainty due to scale variation), $\mu_{\mathcal{B}}^{\text{pf}}$ should be varied accordingly. 

There are five scale variations that we need to consider: the hard scale, $\mu_H$, which extends to all three regions (I, II, and far tail), the global soft scale, $\mu_{ss}$, which extends only to region I and II, the beam scale, $\mu_{\mathcal{B}}^{\text{II}}$, which only applies in region II, the global soft rapidity scale, $\nu_{ss}$, for both regions I and II, and finally the collinear rapidity scale which corresponds to $\nu_{sc}$ in region II, and to $\nu_{\mathcal{B}}$ in region I. In practice we will control these variations through five corresponding parameters, $\epsilon_H$, $\epsilon_{ss}$, $\epsilon_{\mathcal{B}}$, $\zeta_{ss}$, and $\zeta_{cs}$ ($=+1, -1, 0$). The four profile functions and the hard scale are then defined as follows.
\begin{align}
  \mu_H(M;\epsilon_H) = &(2)^{\epsilon_H} M \nn \\
  \mu_{ss}^{\text{pf}}(E_T,M;\epsilon_H,\epsilon_{ss}) = & g \lp (2)^{\epsilon_{ss}}E_T, (2)^{\epsilon_{H}} M \rp \;, \nn \\
  \mu_{\mathcal{B}}^{\text{pf}}(E_T,M,y,r;\epsilon_H,\epsilon_{ss},\epsilon_{\mathcal{B}}) = & \mu_{ss}^{\text{pf}}(E_T,M;\epsilon_H,\epsilon_{ss}) +(2)^{\epsilon_{\mathcal{B}}} M e^{-y} r \nn \\& - g \lp (2)^{\epsilon_{\mathcal{B}}} M e^{-y} r , \mu_{ss}^{\text{pf}}(E_T,M;\epsilon_H,\epsilon_{ss}) \rp \;, \nn \\
  \nu_{ss}^{\text{pf}}(E_T,M;\zeta_{cs},\zeta_{ss}) = & g \lp (2)^{\zeta_{ss}}E_T, (2)^{\zeta_{cs}} M \rp \;, \nn \\
 \nu_{cs}^{\text{pf}}(E_T, r,M;\zeta_{cs})  = & (2)^{\zeta_{cs}} g \lp E_T/ r, M \rp \;,
  \end{align}
where
\begin{equation}
  g(\mu_1,\mu_2) = \frac{\mu_1}{(1 +(\mu_1/\mu_2)^{n})^{1/n}}.
\end{equation}
We choose $n=10$. In Figure~\ref{fig:scales} we illustrate how each of the profile functions changes when we consider the variation for one of the five control parameters: $\epsilon_H$,  $\epsilon_{ss}$, $\epsilon_{\mathcal{B}}$, $\zeta_{ss}$, and $\zeta_{cs}$. In each plot we separate the three different regions with vertical lines.

One can also explore the sensitivity of the differential cross section to the choice of the function $g(\mu_1,\mu_2)$. In our formalism this can be performed by setting all variation control parameters to zero and varying the parameter $n$. We find that for $4 < n < 12$ the result falls within the error bands of the scale variation. We find that $n=10$ gives the best numerical stability for the central values.  We do not explore different parameterizations of the function $g(\mu_1,\mu_2)$.

\bibliography{paper}

\providecommand{\href}[2]{#2}\begingroup\raggedright\begin{thebibliography}{10}

\bibitem{CMS:2016etb}
{\scshape CMS} collaboration, \emph{{Measurement of the underlying event using
  the Drell-Yan process in proton-proton collisions at $\sqrt{s} =
  13~\mathrm{TeV}$}}, {\emph{CMS-PAS-FSQ-16-008} }.

\bibitem{Chatrchyan:2012tb}
{\scshape CMS} collaboration, S.~Chatrchyan et~al., \emph{{Measurement of the
  underlying event in the Drell-Yan process in proton-proton collisions at
  $\sqrt{s}=7$ TeV}},
  \href{http://dx.doi.org/10.1140/epjc/s10052-012-2080-4}{\emph{Eur. Phys. J.}
  {\bf C72} (2012) 2080}, [\href{http://arxiv.org/abs/1204.1411}{{\tt
  1204.1411}}].

\bibitem{Aad:2016ria}
{\scshape ATLAS} collaboration, G.~Aad et~al., \emph{{Measurement of
  event-shape observables in $Z \rightarrow \ell ^{+} \ell ^{-}$ events in $pp$
  collisions at $\sqrt{s}=$ 7 TeV with the ATLAS detector at the LHC}},
  \href{http://dx.doi.org/10.1140/epjc/s10052-016-4176-8}{\emph{Eur. Phys. J.}
  {\bf C76} (2016) 375}, [\href{http://arxiv.org/abs/1602.08980}{{\tt
  1602.08980}}].

\bibitem{Aaboud:2017fwp}
{\scshape ATLAS} collaboration, M.~Aaboud et~al., \emph{{Measurement of
  charged-particle distributions sensitive to the underlying event in $
  \sqrt{s}=13 $ TeV proton-proton collisions with the ATLAS detector at the
  LHC}}, \href{http://dx.doi.org/10.1007/JHEP03(2017)157}{\emph{JHEP} {\bf 03}
  (2017) 157}, [\href{http://arxiv.org/abs/1701.05390}{{\tt 1701.05390}}].

\bibitem{Aad:2014hia}
{\scshape ATLAS} collaboration, G.~Aad et~al., \emph{{Measurement of the
  underlying event in jet events from 7 TeV proton-proton collisions with the
  ATLAS detector}},
  \href{http://dx.doi.org/10.1140/epjc/s10052-014-2965-5}{\emph{Eur. Phys. J.}
  {\bf C74} (2014) 2965}, [\href{http://arxiv.org/abs/1406.0392}{{\tt
  1406.0392}}].

\bibitem{Stewart:2009yx}
I.~W. Stewart, F.~J. Tackmann and W.~J. Waalewijn, \emph{{Factorization at the
  LHC: From PDFs to Initial State Jets}},
  \href{http://dx.doi.org/10.1103/PhysRevD.81.094035}{\emph{Phys. Rev.} {\bf
  D81} (2010) 094035}, [\href{http://arxiv.org/abs/0910.0467}{{\tt
  0910.0467}}].

\bibitem{Stewart:2010pd}
I.~W. Stewart, F.~J. Tackmann and W.~J. Waalewijn, \emph{{The Beam Thrust Cross
  Section for Drell-Yan at NNLL Order}},
  \href{http://dx.doi.org/10.1103/PhysRevLett.106.032001}{\emph{Phys. Rev.
  Lett.} {\bf 106} (2011) 032001}, [\href{http://arxiv.org/abs/1005.4060}{{\tt
  1005.4060}}].

\bibitem{Berger:2010xi}
C.~F. Berger, C.~Marcantonini, I.~W. Stewart, F.~J. Tackmann and W.~J.
  Waalewijn, \emph{{Higgs Production with a Central Jet Veto at NNLL+NNLO}},
  \href{http://dx.doi.org/10.1007/JHEP04(2011)092}{\emph{JHEP} {\bf 04} (2011)
  092}, [\href{http://arxiv.org/abs/1012.4480}{{\tt 1012.4480}}].

\bibitem{Banfi:2010xy}
A.~Banfi, G.~P. Salam and G.~Zanderighi, \emph{{Phenomenology of event shapes
  at hadron colliders}},
  \href{http://dx.doi.org/10.1007/JHEP06(2010)038}{\emph{JHEP} {\bf 06} (2010)
  038}, [\href{http://arxiv.org/abs/1001.4082}{{\tt 1001.4082}}].

\bibitem{Rothstein:2016bsq}
I.~Z. Rothstein and I.~W. Stewart, \emph{{An Effective Field Theory for Forward
  Scattering and Factorization Violation}},
  \href{http://dx.doi.org/10.1007/JHEP08(2016)025}{\emph{JHEP} {\bf 08} (2016)
  025}, [\href{http://arxiv.org/abs/1601.04695}{{\tt 1601.04695}}].

\bibitem{Gaunt:2014ska}
J.~R. Gaunt, \emph{{Glauber Gluons and Multiple Parton Interactions}},
  \href{http://dx.doi.org/10.1007/JHEP07(2014)110}{\emph{JHEP} {\bf 07} (2014)
  110}, [\href{http://arxiv.org/abs/1405.2080}{{\tt 1405.2080}}].

\bibitem{Zeng:2015iba}
M.~Zeng, \emph{{Drell-Yan process with jet vetoes: breaking of generalized
  factorization}}, \href{http://dx.doi.org/10.1007/JHEP10(2015)189}{\emph{JHEP}
  {\bf 10} (2015) 189}, [\href{http://arxiv.org/abs/1507.01652}{{\tt
  1507.01652}}].

\bibitem{Papaefstathiou:2010bw}
A.~Papaefstathiou, J.~M. Smillie and B.~R. Webber, \emph{{Resummation of
  transverse energy in vector boson and Higgs boson production at hadron
  colliders}}, \href{http://dx.doi.org/10.1007/JHEP04(2010)084}{\emph{JHEP}
  {\bf 04} (2010) 084}, [\href{http://arxiv.org/abs/1002.4375}{{\tt
  1002.4375}}].

\bibitem{Grazzini:2014uha}
M.~Grazzini, A.~Papaefstathiou, J.~M. Smillie and B.~R. Webber,
  \emph{{Resummation of the transverse-energy distribution in Higgs boson
  production at the Large Hadron Collider}},
  \href{http://dx.doi.org/10.1007/JHEP09(2014)056}{\emph{JHEP} {\bf 09} (2014)
  056}, [\href{http://arxiv.org/abs/1403.3394}{{\tt 1403.3394}}].

\bibitem{Hornig:2017pud}
A.~Hornig, D.~Kang, Y.~Makris and T.~Mehen, \emph{{Transverse Vetoes with
  Rapidity Cutoff in SCET}},
  \href{http://dx.doi.org/10.1007/JHEP12(2017)043}{\emph{JHEP} {\bf 12} (2017)
  043}, [\href{http://arxiv.org/abs/1708.08467}{{\tt 1708.08467}}].

\bibitem{Bauer:2000ew}
C.~W. Bauer, S.~Fleming and M.~E. Luke, \emph{{Summing Sudakov logarithms in $B
  \to X_s \gamma$ in effective field theory}},
  \href{http://dx.doi.org/10.1103/PhysRevD.63.014006}{\emph{Phys. Rev.} {\bf
  D63} (2000) 014006}, [\href{http://arxiv.org/abs/hep-ph/0005275}{{\tt
  hep-ph/0005275}}].

\bibitem{Bauer:2000yr}
C.~W. Bauer, S.~Fleming, D.~Pirjol and I.~W. Stewart, \emph{{An Effective field
  theory for collinear and soft gluons: Heavy to light decays}},
  \href{http://dx.doi.org/10.1103/PhysRevD.63.114020}{\emph{Phys. Rev.} {\bf
  D63} (2001) 114020}, [\href{http://arxiv.org/abs/hep-ph/0011336}{{\tt
  hep-ph/0011336}}].

\bibitem{Bauer:2001ct}
C.~W. Bauer and I.~W. Stewart, \emph{{Invariant operators in collinear
  effective theory}},
  \href{http://dx.doi.org/10.1016/S0370-2693(01)00902-9}{\emph{Phys. Lett.}
  {\bf B516} (2001) 134--142}, [\href{http://arxiv.org/abs/hep-ph/0107001}{{\tt
  hep-ph/0107001}}].

\bibitem{Bauer:2001yt}
C.~W. Bauer, D.~Pirjol and I.~W. Stewart, \emph{{Soft collinear factorization
  in effective field theory}},
  \href{http://dx.doi.org/10.1103/PhysRevD.65.054022}{\emph{Phys. Rev.} {\bf
  D65} (2002) 054022}, [\href{http://arxiv.org/abs/hep-ph/0109045}{{\tt
  hep-ph/0109045}}].

\bibitem{Tackmann:2012bt}
F.~J. Tackmann, J.~R. Walsh and S.~Zuberi, \emph{{Resummation Properties of Jet
  Vetoes at the LHC}},
  \href{http://dx.doi.org/10.1103/PhysRevD.86.053011}{\emph{Phys. Rev.} {\bf
  D86} (2012) 053011}, [\href{http://arxiv.org/abs/1206.4312}{{\tt
  1206.4312}}].

\bibitem{Stewart:2010qs}
I.~W. Stewart, F.~J. Tackmann and W.~J. Waalewijn, \emph{{The Quark Beam
  Function at NNLL}},
  \href{http://dx.doi.org/10.1007/JHEP09(2010)005}{\emph{JHEP} {\bf 09} (2010)
  005}, [\href{http://arxiv.org/abs/1002.2213}{{\tt 1002.2213}}].

\bibitem{Fleming:2006cd}
S.~Fleming, A.~K. Leibovich and T.~Mehen, \emph{{Resummation of Large Endpoint
  Corrections to Color-Octet $J/\psi$ Photoproduction}},
  \href{http://dx.doi.org/10.1103/PhysRevD.74.114004}{\emph{Phys. Rev.} {\bf
  D74} (2006) 114004}, [\href{http://arxiv.org/abs/hep-ph/0607121}{{\tt
  hep-ph/0607121}}].

\bibitem{Chiu:2012ir}
J.-Y. Chiu, A.~Jain, D.~Neill and I.~Z. Rothstein, \emph{{A Formalism for the
  Systematic Treatment of Rapidity Logarithms in Quantum Field Theory}},
  \href{http://dx.doi.org/10.1007/JHEP05(2012)084}{\emph{JHEP} {\bf 1205}
  (2012) 084}, [\href{http://arxiv.org/abs/1202.0814}{{\tt 1202.0814}}].

\bibitem{Chiu:2011qc}
J.-y. Chiu, A.~Jain, D.~Neill and I.~Z. Rothstein, \emph{{The Rapidity
  Renormalization Group}},
  \href{http://dx.doi.org/10.1103/PhysRevLett.108.151601}{\emph{Phys. Rev.
  Lett.} {\bf 108} (2012) 151601}, [\href{http://arxiv.org/abs/1104.0881}{{\tt
  1104.0881}}].

\bibitem{Chien:2015cka}
Y.-T. Chien, A.~Hornig and C.~Lee, \emph{{Soft-collinear mode for jet cross
  sections in soft collinear effective theory}},
  \href{http://dx.doi.org/10.1103/PhysRevD.93.014033}{\emph{Phys. Rev.} {\bf
  D93} (2016) 014033}, [\href{http://arxiv.org/abs/1509.04287}{{\tt
  1509.04287}}].

\bibitem{Bauer:2011uc}
C.~W. Bauer, F.~J. Tackmann, J.~R. Walsh and S.~Zuberi, \emph{{Factorization
  and Resummation for Dijet Invariant Mass Spectra}},
  \href{http://dx.doi.org/10.1103/PhysRevD.85.074006}{\emph{Phys. Rev.} {\bf
  D85} (2012) 074006}, [\href{http://arxiv.org/abs/1106.6047}{{\tt
  1106.6047}}].

\bibitem{Alwall:2014hca}
J.~Alwall, R.~Frederix, S.~Frixione, V.~Hirschi, F.~Maltoni, O.~Mattelaer
  et~al., \emph{{The automated computation of tree-level and next-to-leading
  order differential cross sections, and their matching to parton shower
  simulations}}, \href{http://dx.doi.org/10.1007/JHEP07(2014)079}{\emph{JHEP}
  {\bf 07} (2014) 079}, [\href{http://arxiv.org/abs/1405.0301}{{\tt
  1405.0301}}].

\bibitem{Sjostrand:2006za}
T.~Sjostrand, S.~Mrenna and P.~Z. Skands, \emph{{PYTHIA 6.4 Physics and
  Manual}}, \href{http://dx.doi.org/10.1088/1126-6708/2006/05/026}{\emph{JHEP}
  {\bf 05} (2006) 026}, [\href{http://arxiv.org/abs/hep-ph/0603175}{{\tt
  hep-ph/0603175}}].

\bibitem{Sjostrand:2007gs}
T.~Sjostrand, S.~Mrenna and P.~Z. Skands, \emph{{A Brief Introduction to PYTHIA
  8.1}}, \href{http://dx.doi.org/10.1016/j.cpc.2008.01.036}{\emph{Comput. Phys.
  Commun.} {\bf 178} (2008) 852--867},
  [\href{http://arxiv.org/abs/0710.3820}{{\tt 0710.3820}}].

\bibitem{Korchemsky:2000kp}
G.~P. Korchemsky and S.~Tafat, \emph{{On power corrections to the event shape
  distributions in QCD}},
  \href{http://dx.doi.org/10.1088/1126-6708/2000/10/010}{\emph{JHEP} {\bf 10}
  (2000) 010}, [\href{http://arxiv.org/abs/hep-ph/0007005}{{\tt
  hep-ph/0007005}}].

\bibitem{Bauer:2002ie}
C.~W. Bauer, A.~V. Manohar and M.~B. Wise, \emph{{Enhanced nonperturbative
  effects in jet distributions}},
  \href{http://dx.doi.org/10.1103/PhysRevLett.91.122001}{\emph{Phys. Rev.
  Lett.} {\bf 91} (2003) 122001},
  [\href{http://arxiv.org/abs/hep-ph/0212255}{{\tt hep-ph/0212255}}].

\bibitem{Lee:2006fn}
C.~Lee and G.~F. Sterman, \emph{{Universality of nonperturbative effects in
  event shapes}}, {\emph{eConf} {\bf C0601121} (2006) A001},
  [\href{http://arxiv.org/abs/hep-ph/0603066}{{\tt hep-ph/0603066}}].

\bibitem{Lee:2007jr}
C.~Lee, \emph{{Universal nonperturbative effects in event shapes from
  soft-collinear effective theory}},
  \href{http://dx.doi.org/10.1142/S021773230702289X}{\emph{Mod. Phys. Lett.}
  {\bf A22} (2007) 835--851}, [\href{http://arxiv.org/abs/hep-ph/0703030}{{\tt
  hep-ph/0703030}}].

\bibitem{Hornig:2009vb}
A.~Hornig, C.~Lee and G.~Ovanesyan, \emph{Effective predictions of event
  shapes: Factorized, resummed, and gapped angularity distributions},
  \href{http://dx.doi.org/10.1088/1126-6708/2009/05/122}{\emph{JHEP} {\bf 05}
  (2009) 122}, [\href{http://arxiv.org/abs/0901.3780}{{\tt 0901.3780}}].

\bibitem{Kang:2013lga}
Z.-B. Kang, X.~Liu and S.~Mantry, \emph{{1-jettiness DIS event shape: NNLL+NLO
  results}}, \href{http://dx.doi.org/10.1103/PhysRevD.90.014041}{\emph{Phys.
  Rev.} {\bf D90} (2014) 014041}, [\href{http://arxiv.org/abs/1312.0301}{{\tt
  1312.0301}}].

\bibitem{Kang:2018qra}
Z.-B. Kang, K.~Lee and F.~Ringer, \emph{{Jet angularity measurements for single
  inclusive jet production}},
  \href{http://dx.doi.org/10.1007/JHEP04(2018)110}{\emph{JHEP} {\bf 04} (2018)
  110}, [\href{http://arxiv.org/abs/1801.00790}{{\tt 1801.00790}}].

\bibitem{Moult:2017okx}
I.~Moult, B.~Nachman and D.~Neill, \emph{{Convolved Substructure: Analytically
  Decorrelating Jet Substructure Observables}},
  \href{http://dx.doi.org/10.1007/JHEP05(2018)002}{\emph{JHEP} {\bf 05} (2018)
  002}, [\href{http://arxiv.org/abs/1710.06859}{{\tt 1710.06859}}].

\bibitem{Stewart:2014nna}
I.~W. Stewart, F.~J. Tackmann and W.~J. Waalewijn, \emph{{Dissecting Soft
  Radiation with Factorization}},
  \href{http://dx.doi.org/10.1103/PhysRevLett.114.092001}{\emph{Phys. Rev.
  Lett.} {\bf 114} (2015) 092001}, [\href{http://arxiv.org/abs/1405.6722}{{\tt
  1405.6722}}].

\bibitem{Becher:2013iya}
T.~Becher and G.~Bell, \emph{{Enhanced nonperturbative effects through the
  collinear anomaly}},
  \href{http://dx.doi.org/10.1103/PhysRevLett.112.182002}{\emph{Phys. Rev.
  Lett.} {\bf 112} (2014) 182002}, [\href{http://arxiv.org/abs/1312.5327}{{\tt
  1312.5327}}].

\bibitem{Hoang:2017kmk}
A.~H. Hoang, S.~Mantry, A.~Pathak and I.~W. Stewart, \emph{{Extracting a Short
  Distance Top Mass with Light Grooming}},
  \href{http://arxiv.org/abs/1708.02586}{{\tt 1708.02586}}.

\bibitem{Kang:2018jwa}
Z.-B. Kang, K.~Lee, X.~Liu and F.~Ringer, \emph{{The groomed and ungroomed jet
  mass distribution for inclusive jet production at the LHC}},
  \href{http://arxiv.org/abs/1803.03645}{{\tt 1803.03645}}.

\bibitem{Bahr:2008pv}
M.~Bahr et~al., \emph{{Herwig++ Physics and Manual}},
  \href{http://dx.doi.org/10.1140/epjc/s10052-008-0798-9}{\emph{Eur. Phys. J.}
  {\bf C58} (2008) 639--707}, [\href{http://arxiv.org/abs/0803.0883}{{\tt
  0803.0883}}].

\bibitem{Ritzmann:2014mka}
M.~Ritzmann and W.~J. Waalewijn, \emph{{Fragmentation in Jets at NNLO}},
  \href{http://dx.doi.org/10.1103/PhysRevD.90.054029}{\emph{Phys.Rev.} {\bf
  D90} (2014) 054029}, [\href{http://arxiv.org/abs/1407.3272}{{\tt
  1407.3272}}].

\bibitem{Altarelli:1977zs}
G.~Altarelli and G.~Parisi, \emph{{Asymptotic Freedom in Parton Language}},
  \href{http://dx.doi.org/10.1016/0550-3213(77)90384-4}{\emph{Nucl. Phys.} {\bf
  B126} (1977) 298--318}.

\bibitem{PhysRevD.9.980}
D.~J. Gross and F.~Wilczek, \emph{Asymptotically free gauge theories. ii},
  \href{http://dx.doi.org/10.1103/PhysRevD.9.980}{\emph{Phys. Rev. D} {\bf 9}
  (Feb, 1974) 980--993}.

\end{thebibliography}\endgroup

\end{document}